\documentclass[twocolumn]{aastex61}
\usepackage{graphicx}

\usepackage{natbib}
\bibpunct{(}{)}{;}{a}{}{,} % to follow the A&A style
\usepackage{amstext}
\usepackage{url}
\usepackage{subfigure}
\usepackage{mathtools}
\usepackage[T1]{fontenc}
\shorttitle{G182.5--4.0 and the coherent magnetic field in the outer Galaxy}
\shortauthors{West et al.}
\accepted{15 October 2022}
\submitjournal{ApJ}

\begin{document}

\title{Discovery of a filamentary synchrotron structure connected to the coherent magnetic field in the outer Galaxy}

\correspondingauthor{J. L. West et al.}
\email{jennifer.west@dunlap.utoronto.ca}

\author[0000-0001-7722-8458]{J. L. West}
\affil{Dunlap Institute for Astronomy and Astrophysics, University of Toronto, Toronto, M5S 3H4, Canada}

\author{J. L. Campbell}
\affil{Dunlap Institute for Astronomy and Astrophysics, University of Toronto, Toronto, M5S 3H4, Canada}
\affil{David A. Dunlap Department of Astronomy and Astrophysics, University of Toronto, Toronto, M5S 3H4, Canada}

\author{P. Bhaura}
\affil{David A. Dunlap Department of Astronomy and Astrophysics, University of Toronto, Toronto, M5S 3H4, Canada}

\author{R. Kothes}
\affil{National Research Council Canada, Herzberg Research Centre for Astronomy and Astrophysics, Dominion Radio Astrophysical Observatory, PO Box 248, Penticton, V2A 6J9, Canada}

\author{S. Safi-Harb}
\affil{Department of Physics and Astronomy, University of Manitoba,  Winnipeg, R3T 2N2, Canada}

\author{J. M. Stil}
\affil{Department of Physics and Astronomy, The University of Calgary, 2500 University Drive NW, Calgary AB T2N 1N4, Canada}

\author{A. R. Taylor}
\affil{Inter-Unversity Institute for Data Intensive Astronomy}
\affil{University of Cape Town, South Africa}
\affil{University of the Western Cape, South Africa}

\author{T. Foster}
\affil{Department of Physics and Astronomy, Brandon University, 270 - 18th Street, Brandon, R7A 6A9, Canada}

\author[0000-0002-3382-9558]{B. M. Gaensler}
\affil{Dunlap Institute for Astronomy and Astrophysics, University of Toronto, Toronto, ON M5S 3H4, Canada}
\affil{David A. Dunlap Department of Astronomy and Astrophysics, University of Toronto, Toronto, ON M5S 3H4, Canada}

\author{S. J. George}
\affil{School of Physics and Astronomy, University of Birmingham, UK, B15 2TT}

\author[0000-0002-1495-760X]{S. J. Gibson}
\affil{Department of Physics and Astronomy, Western Kentucky University, Bowling Green, KY 42101 USA}

\author{R. Ricci}
\affil{Instituto Nazionale Ricerche Metrologiche - Strada delle Cacce 91, Torino, 10135 Italy}
\affil{INAF-Istituto di Radioastronomia, Via Gobetti 101, Bologna, 40129 Italy}

\begin{abstract}

Using data from the Galactic Arecibo L-band Feed Array Continuum Transit Survey (GALFACTS), we report the discovery of two previously unidentified, very compressed, thin, and straight polarized filaments approximately centred at Galactic coordinates, $(l,b)=(182.5^\circ,-4.0^\circ)$, which we call G182.5--4.0. Using data from the Isaac Newton Telescope Galactic Plane Survey (IGAPS), we also find straight, long, and extremely thin H$\alpha$ filaments coincident with the radio emission.  These filaments are positioned in projection at the edge of the Orion-Eridanus superbubble and we find evidence indicating that the filaments align with the coherent magnetic field of the outer Galaxy. We find a lower limit on the total radio flux at 1.4~GHz to be $0.7\pm0.3$~Jy with an average linearly polarized fraction of $40\substack{+30 \\ -20}\%$.  We consider various scenarios that could explain the origin of these filaments, including a shell-type supernova remnant (SNR), a bow shock nebula associated with a pulsar, or relic fragments from one or more supernova explosions in the adjacent superbubble, with a hybrid scenario being most likely. This may represent an example of a new class of objects that is neither an SNR nor a bow shock. The highly compressed nature of these filaments and their alignment with Galactic plane suggests a scenario where this object formed in a magnetic field that was compressed by the expanding Orion-Eridanus superbubble, suggesting that the object is related to this superbubble and implying a distance of $\sim$400~pc.

\end{abstract}

\section{\label{sec:intro}Introduction}

As the resolution and sensitivity of observations improve, we are increasingly finding that the interstellar medium (ISM) of our Galaxy is filled with filamentary (i.e., long and narrow) structures from a variety of different origins. These filaments are observed in emission across the electromagnetic spectrum including with X-rays \citep[e.g.,][]{2022HEAD...1911008D}, ultraviolet \citep[e.g.,][]{2020A&A...636L...8B}, optical \citep[e.g.,][]{2021ApJ...920...90F}, infrared \citep[e.g.,][]{2022arXiv220306331G}, and radio \citep[e.g.,][]{1984Natur.310..557Y, 2018MNSSA..77..102.}, and there are also long and narrow structures detected through non-emitting tracers such as Faraday rotation \citep[e.g.,][]{2022ApJ...927...49C} and scintillation of background sources \citep[e.g.,][]{2021MNRAS.502.3294W}. The filaments are associated with many different environments and origins including star forming regions, supernova remnants (SNRs), bow shock nebulae, and also filaments of uncertain origins \citep[e.g.,][]{2015A&A...583A.137J,2020A&A...636L...8B}, and the radiation can come from both thermal and non-thermal (i.e., synchrotron) emission mechanisms. 

There is also evidence that the orientation of filaments are connected to magnetic fields  \citep{2014ApJ...789...82C, 2015MNRAS.454L..46Z, 2015A&A...583A.137J, 2016A&A...586A.135P, 2021ApJ...923...58W, 2022ApJ...927...49C}. Spiral galaxies, including the Milky Way, are known to have coherent large (galactic) scale magnetic fields \citep{2015A&ARv..24....4B}, which are thought originate through the $\alpha-\Omega$ dynamo and compression from spiral shocks and supernova (SN) explosions \citep{2015A&ARv..24....4B}. Outflows, perhaps from superbubbles, driven by SNe are needed to drive the dynamo process \citep{Parker:1992iz}. 

There is strong evidence that the  shells of SNRs align with the large-scale, mean Galactic magnetic field on scales $\gtrsim10$~pc \citep{1977PASA....3..130C,1998ApJ...493..781G,2016A&A...587A.148W}. The blast wave of the SN compresses the ambient medium, freezing in the magnetic field. As the SNR ages, it is thought to become stretched out and elongated parallel to the magnetic field. The timeline depends strongly on the magnetic tension and therefore the magnetic field strength. In the Galactic centre, there are many straight and compressed non-thermal filaments \citep{1984Natur.310..557Y, 2018MNSSA..77..102.} that are thought to possibly be relics of old SNRs \citep{2020PASJ..tmp..161S}. There is also increasing evidence that the neutral hydrogen (HI) component of the Galaxy contains filamentary structures, or ``fibres'', that align with the Galactic magnetic field \citep{2014ApJ...789...82C}.

\begin{figure*}[!ht]
\centering \includegraphics[width=18cm]{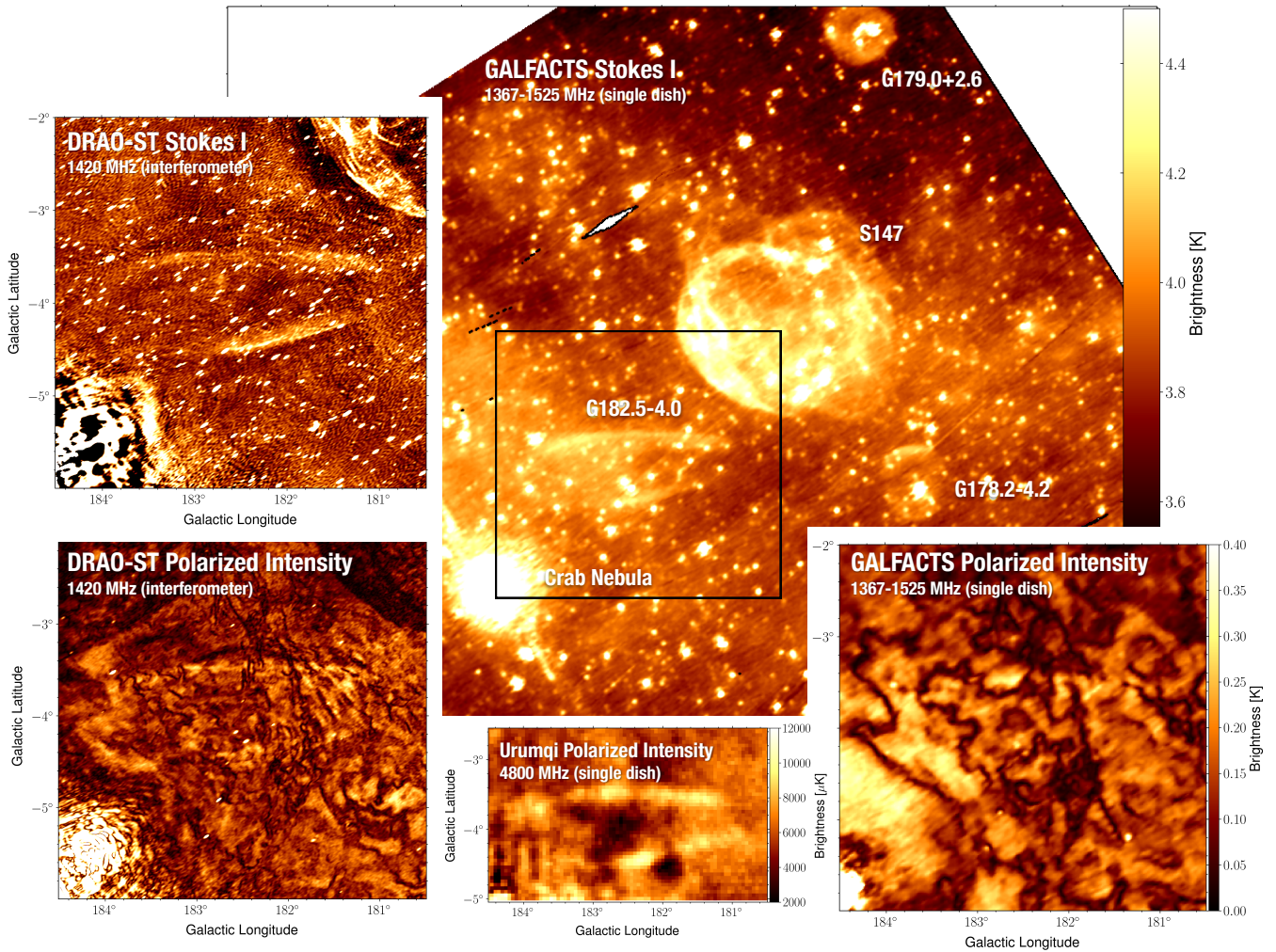}
\caption{\label{fig:continuum}G182.5 and surrounding area for the GALFACTS Stokes $I$ (main image), DRAO-ST Stokes $I$ (top left inset), DRAO-ST  polarized intensity (bottom left inset), Urumqi polarized intensity data (bottom centre inset) and GALFACTS peak polarized intensity data (bottom right inset).  The black square in the main image indicates the region shown in the inset images. Prominent features in the surrounding area are labeled. 
}
\end{figure*}

\begin{figure*}[!ht]
\centering \includegraphics[width=18cm]{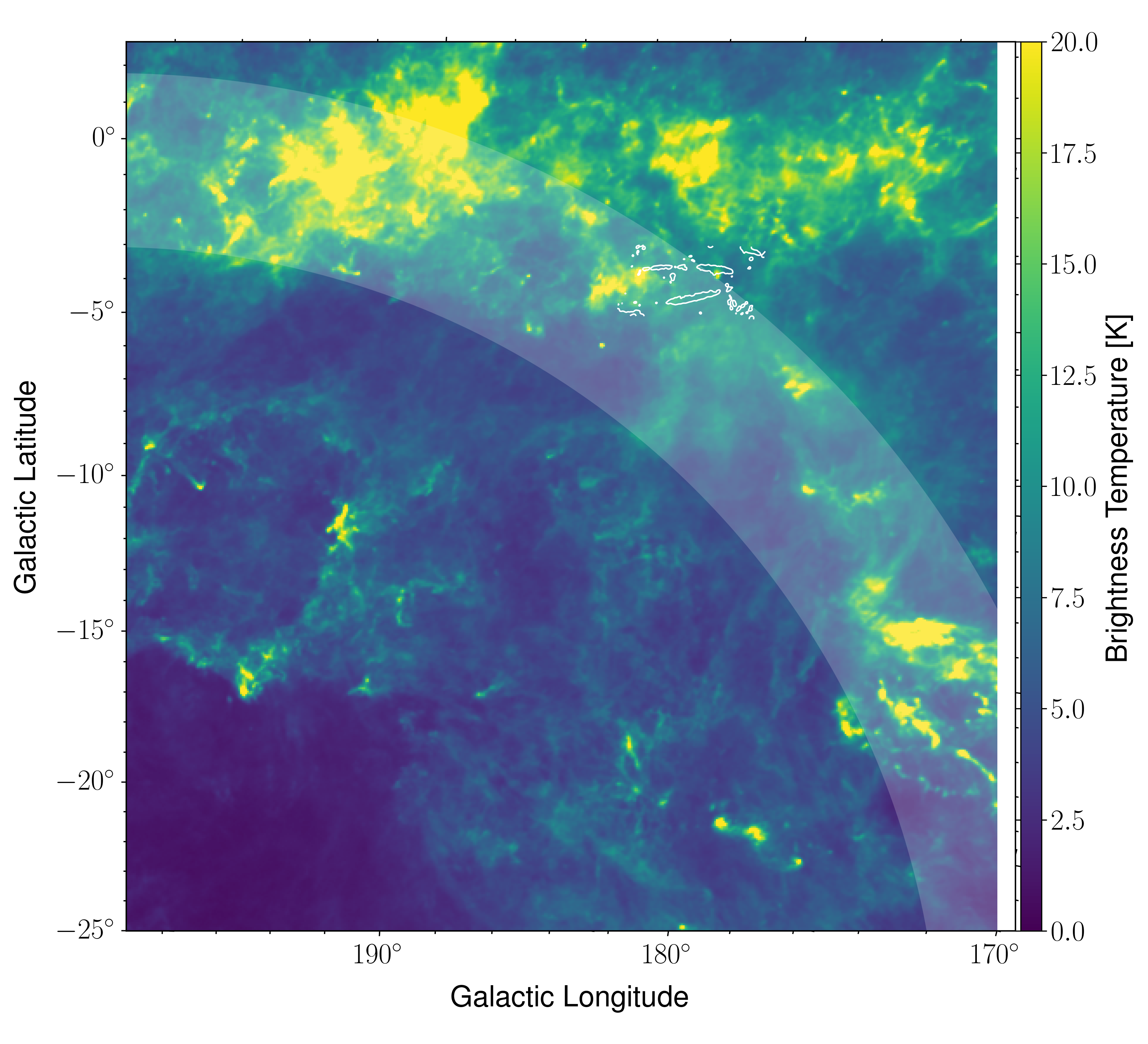}

\begin{scriptsize}\caption{\label{fig:orioneridanus}\textit{Planck} 545 GHz dust image showing a portion of the Orion-Eridanus superbubble. The approximate edge of the bubble is shown by the shaded white band \citep[based on][]{2015ApJ...808..111O, 2022A&A...660L...7T}. The white contours near the top-right are the DRAO Stokes $I$ data showing the position and orientation of the G182.5--4.0 filaments.
}
\end{scriptsize} 
\end{figure*}

We have discovered two previously unidentified, faint, radio synchrotron emitting
filaments located towards the outer Galaxy, shown in Fig.~\ref{fig:continuum}. We assume that these filaments have the same origin, and thus we consider them to be part of a single structure, which we call G182.5--4.0. The filaments are located at the edge of the Orion-Eridanus superbubble \citep[distance $\approx400$~pc,][]{2015ApJ...808..111O}, as shown in Fig.~\ref{fig:orioneridanus}. The structure shares similarities both with known SNRs and bow shock nebulae, but yet it does not conform to the typical properties of either of these classes of object. 

Radio observations of non-thermal synchrotron radiation implies the existence of relativistic electrons in a magnetic field. This radiation is intrinsically polarized with a negative spectral index, $\alpha<0$ (where the flux density, $S_{\nu}\propto\nu^{\alpha}$), which distinguishes it from thermal bremsstrahlung radiation that can also be bright at radio frequencies. Radio synchrotron brightness is proportional to the plane-of-sky magnetic field component, $B_{\perp}$. Polarized radiation also undergoes Faraday rotation, the study of which reveals additional information about the line-of-sight (LOS) magnetic field strength ($B_{\parallel}$) and direction. 

An electromagnetic wave undergoes Faraday rotation as it propagates through a magneto-ionic medium. The Faraday depth, $\phi(d)$, is defined as the integral of the magneto-ionic medium along a LOS, from a distance, $r=d$, to the observer at $r=0$,

\begin{equation}
\centering
\phi(d) = 0.812\int\limits_0^d {{n_e}{B_\parallel }} dr {\rm{[rad/}}{{\rm{m}}^{\rm{2}}}{\rm{]}}.
\end{equation}
Here $n_e$ [cm$^{-3}$] is the thermal electron density and ${B_\parallel>0}$ [$\mu$G] when the magnetic field is pointed towards the observer. 

In the simplest observational case, $\phi(d)$ can be found by measuring the dependence of the position angle of the polarization vector, $\chi$, as a function of $\lambda^2$ [m$^2]$. When the total Faraday depth of a background source is measured through a Faraday-rotating medium, then we call this the Faraday rotation measure, RM, where

\begin{equation}
\label{eqn:chiobs}
\centering
\chi_\text{obs}=\chi_\text{src}+\text{RM}\lambda^2.
\end{equation}
The amount of rotation (i.e., the difference between $\chi_{obs}$ and $\chi_{src}$) is $\propto\lambda^2$, and thus can be significantly greater at longer wavelengths.

In the case of diffuse synchrotron emission, where there is both rotation and emission along the same LOS, %the medium is said to be Faraday thick, and 
the amount of rotation (and thus the Faraday depth) will vary for each point along the LOS where the emission is taking place. Here, the simple linear relationship between $\chi$ and $\lambda^2$ no longer holds and instead we use the technique of RM-synthesis \citep{Brentjens2005} to measure the polarized intensity as a function of Faraday depth where

\begin{equation}
\centering
{\bf{F}}(\phi ) = \int\limits_{ - \infty }^\infty  {{\bf{P}}({\lambda ^2}){e^{ - 2i\phi {\lambda ^2}}}d{\lambda ^2}}.
\end{equation}
Here, ${\bf{P}} ({\lambda ^2})=Q({\lambda ^2}) +iU({\lambda ^2})$ is the linearly polarized intensity, where $Q$ and $U$, are the linear polarization Stokes parameters, and $\bf{F}(\phi )$ is the polarized intensity as a function of Faraday depth. From this process, we measure a Faraday depth spectrum. The simplest analysis measures the peak of this spectrum, $\phi_\text{peak}$, which tells us the average Faraday depth integrated along the LOS.

\begin{figure*}[!ht]
\centering \includegraphics[width=18cm]{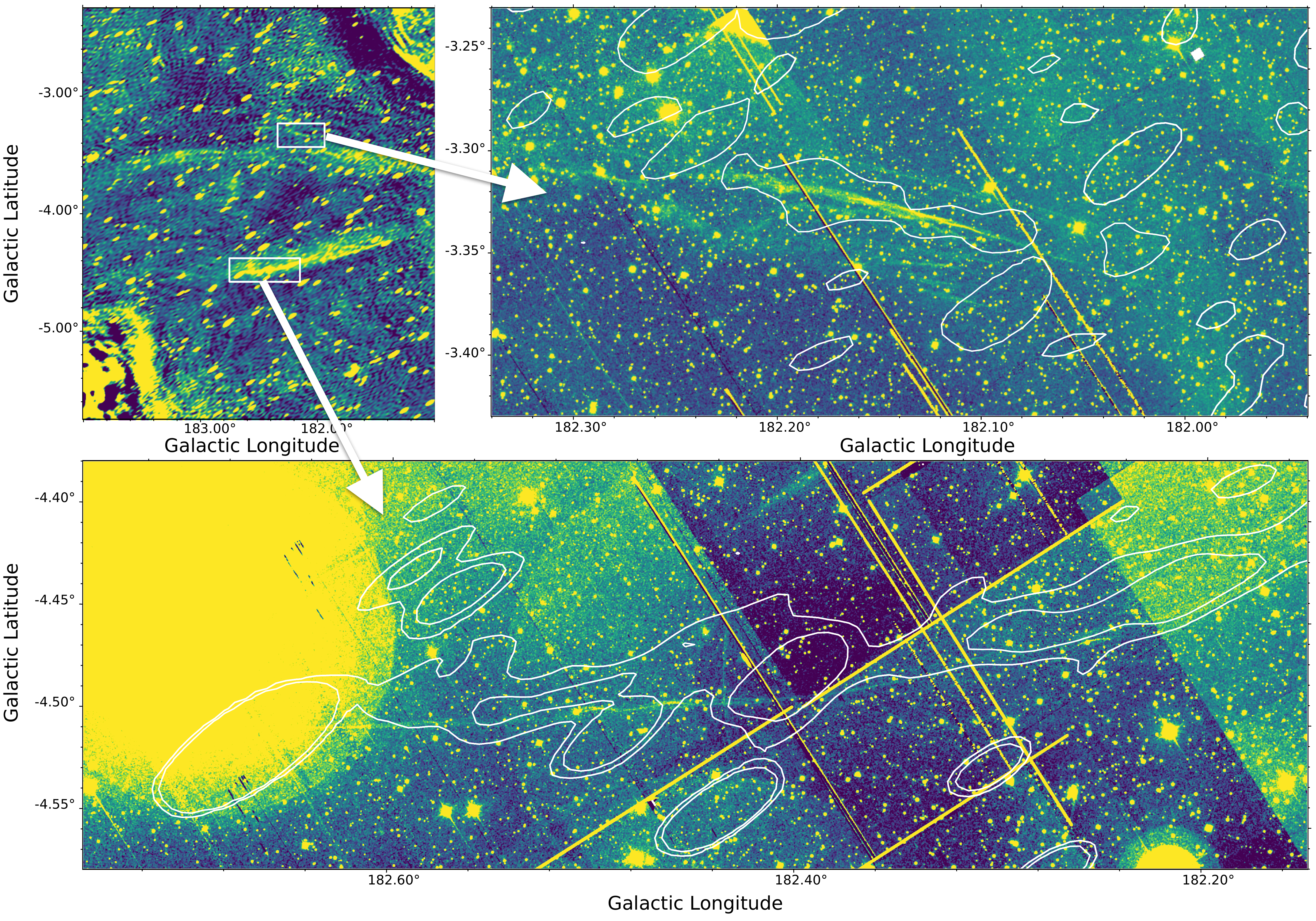}
\caption{\label{fig:iphas}H-$\alpha$ filaments detected in the IGAPS data with contours from Stokes $I$. The DRAO Stokes $I$ image is shown at upper left with white rectangles indicating the regions shown in the H-$\alpha$ mosaics. In all panels, yellow show bright emission and dark blue is faint.
}
\end{figure*}

In this paper, we explore the properties of G182.5--4.0 and its possible connection to the large-scale Galactic magnetic field with the goal of understanding its origin. In Sec.~\ref{sec:observations} we describe the data that we use to measure the structure's properties, which are presented in Sec.~\ref{sec:properties}. We present a radial profile (Sec.~\ref{sec:profile}) and look for evidence connecting these filaments to the large-scale Galactic field using starlight polarization (Sec.~\ref{sec:polstars}), neutral hydrogen structures (Sec.~\ref{sec:rht}), and Faraday rotation (Sec.~\ref{sec:rmsynth}). Discussion and conclusions are presented in Sec.~\ref{sec:discussion} and Sec.~\ref{sec:conclusions}, respectively.

\section{\label{sec:observations}Observations}

\subsection{\label{sec:arecibo}Arecibo}
G182.5--4.0 consists of two straight and nearly parallel filaments, first identified in data from the Galactic Arecibo L-band Feed Array Continuum Transit Survey (GALFACTS) using the 305~m Arecibo radio telescope (data release version 3.1.2). These data have full Stokes polarization ($I$, $Q$, $U$, and $V$) observed over 1367--1525~MHz, with a channel width of 0.42~MHz \citep{2010ASPC..438..402T}. These data have a beam FWHM$\approx3.5'$ and a band-averaged sensitivity of 90 $\mu$Jy/beam. In Fig.~\ref{fig:continuum} we show the total intensity and peak polarized intensity, $\bf{F}(\phi_\text{peak})$, maps for these data (see Sec.~\ref{sec:rmsynth} for further details).

\subsection{\label{sec:draost}Dominion Radio Astrophysical Observatory Synthesis Telescope}

We obtained observations using the Dominion Radio Astrophysical Observatory Synthesis Telescope \citep[DRAO-ST,][]{2000A&AS..145..509L} with continuum observations at 1.4~GHz in Stokes $I$, $Q$, and $U$, in addition to HI spectral line observations. The main characteristics of this survey are the same as for the Canadian Galactic Plane Survey \citep{cgps}. As we are mapping close to one of the brightest radio sources in the sky, the Crab Nebula, we placed the three fields covering G182.4$-$4.0 so that the Crab Nebula was in the first null of the antenna's primary beam at 1420~MHz,
in order to reduce artifacts.

The spectral line observations cover a velocity range of 256 channels from 85.1~km~s$^{-1}$ to --125.1~km~s$^{-1}$ with a channel width of 0.825~km~s$^{-1}$. The angular resolution varies slightly across the maps as cosec(Declination). At the centre of G182.5$-$4.0 the resolution is 125"$\times$49" at 1420~MHz in Stokes $I$ and 134"$\times$59" in the HI line and in the linear polarization data. The rms noise is about 20~mK or 200$\mu$Jy/beam in the continuum images and about 1~K per 0.825~km~s$^{-1}$ channel for the HI data. These data do not have short spacings added and therefore they are missing the zero level flux.

\subsection{\label{sec:urumqi}Urumqi}
We also use higher frequency (4800 MHz), but lower resolution data (FWHM=$9.5'$) from the Urumqi 25~m telescope \citep{2007A&A...463..993S}\footnote{Data obtained from \url{https://www3.mpifr-bonn.mpg.de/survey.html}}. The relatively high frequency means that the Faraday rotation is negligible for these data, making them useful to derive the polarized fraction. These data are absolute-level corrected with extrapolated WMAP K-band data in IAU convention from the 9-years' release.

\subsection{\label{sec:other}Other data}

We use the combined H$\alpha$ data from the Virginia Tech Spectral line Survey (VTSS) and the Wisconsin H$\alpha$ Mapper \citep[WHAM,][FWHM=6$'$]{2003ApJS..149..405H}\footnote{Data obtained from \url{http://skyview.gsfc.nasa.gov}}, to look for diffuse ionized gas around G182.5$-$4.0. 

In addition, we use the high resolution (FWHM$\sim1\arcsec$) Isaac Newton Telescope (INT) Galactic Plane Survey (IGAPS) to search H$\alpha$, ultraviolet (UV), and other optical data (g, r, and i filters)\footnote{Data obtained from \url{https://www.igapsimages.org}}. 
We identify a multiple long ($>0.5^\circ)$, thin ($<0.0015^\circ)$, and straight H$\alpha$ counterpart to the radio filaments in both the north and the south, shown in Fig.~\ref{fig:iphas}. There is no evidence for these filaments in the other filters, including UV. Other authors \citep[][]{2020A&A...636L...8B, 2021ApJ...920...90F} have detected thin  filaments with H$\alpha$ counterparts in Far-UV using the Galaxy Explorer All Sky Survey (GALEX) data. Unfortunately Far-UV GALEX data for the region around G182.5$-$4.0 is not available so we are unable to check for the presence of these filaments in that filter.

We also use archival data of thermal dust emission from \textit{Planck} at 545~GHz\footnote{Data obtained from \url{http://skyview.gsfc.nasa.gov}} in order to compare the position of G182.5$-$4.0 to the wall of the Orion-Eridanus shell. 

\section{\label{sec:properties}Observed Properties}
%\subsection{\label{sec:properties}}

The two radio filaments are approximately $2.7^\circ$ and $1.7^\circ$ degrees long, with a separation that varies from about $0.5^\circ$ on the Western side to $1.0^\circ$ on the Eastern side (see Fig.~\ref{fig:continuum}). They are oriented roughly parallel to the Galactic plane, with a by-eye orientation of 89$^\circ\pm3^\circ$ (Northern filament), and 104$^\circ\pm3^\circ$ (Southern filament), where 90$^\circ$ is parallel to the Galactic plane. They have an approximate geometric centre point at $(l,b)=(182.5^\circ,-4.0^\circ)$. In the brightest region in the South, the surface brightness is only about 0.2~K above the background level.

We detect long and thin H$\alpha$ filaments coincident with the radio emission (see Fig.~\ref{fig:iphas}) with an aspect ratio of $\approx500$:1. The Southern  H$\alpha$ filament is detected across about $0.5^\circ$ of sky with a width of only $\sim4\arcsec$. They are also very faint, being only 1-2$\sigma$ above the noise limit of the data, which has a magnitude limit of 21 mag \citep{2005MNRAS.362..753D}. Using the IPHAS flux conversion from \citet[][in-band H$\alpha$ flux for Vega is $1.52\times10^{-7}$ erg~cm$^{-2}$~s$^{-1}$]{2014MNRAS.444.3230B}, we find that the filaments have an approximate surface brightness of 1-3~Rayleighs (R) (where 1~R$=5.661\times10^{-18}$~erg~s$^{-1}$~cm$^{-2}$ arcsec$^{-2}$ for H$\alpha$). This is somewhat brighter than the H$\alpha$ counterpart to the 30$^\circ$ long UV filament detected by \citet[][]{2020A&A...636L...8B}, where the brightness was $\approx0.5$~R. Our values are consistent with the brightnesses of evolved SNR filaments found by \citet[][]{2021ApJ...920...90F}. 

Due to uncertainty in the zero point of the GALFACTS map, we use the DRAO Stokes $I$ map to estimate the flux density. We first remove the point sources by fitting and subtracting elliptical gaussians to the unresolved sources. The SNR flux density is then found by taking the sum in a rectangular box measuring $2.75^\circ$ by $1.2^\circ$ centred on $(l,b)=(182.3^\circ,-4.0^\circ)$. We estimate a constant background level and its uncertainty by finding the average and standard deviation of ten $0.5^\circ$ square regions in the region around the main box. We find a flux density $S_{1.4~\text{GHz}}=0.7\pm0.3$~Jy. However, since the DRAO map is missing flux from the absence of zero-spacings, we note that this is only a lower limit on the flux density.

From the DRAO polarized intensity ($P=\sqrt{Q^2+U^2}$) image (bottom left of Fig.~\ref{fig:continuum}), the filaments are distinctly visible. Polarized synchrotron emission has a theoretical maximum polarized fraction of about $70\%$. Observed polarized fraction values are rarely this high due to various depolarizing effects. Depolarization occurs due to changes in the orientation of the polarization pseudo-vector along the line of sight (depth depolarization), within the beam of the telescope (beam depolarization), and across the bandwidth of the observation (bandwidth depolarization). The large GALFACTS beam causes significant beam depolarization in the polarized intensity image  (bottom right of Fig.~\ref{fig:continuum}), making the filaments difficult to see.

Both depth and beam depolarization can occur due to random (turbulent) fluctuations, and they will have much less impact in a highly ordered magnetic field. Faraday rotation can further scramble the polarization pseudo-vectors, enhancing the depolarizing effects of all three mechanisms. In order to decouple the depolarizing effects of Faraday rotation and turbulence, we measure the polarization fraction, $p=P/I$, at high frequencies where the Faraday rotation is negligible. Therefore we use the higher frequency data from the Urumqi telescope (shown in Fig.~\ref{fig:continuum}), which avoids the depolarizing effects of Faraday rotation that impacts the lower frequency DRAO and GALFACTS data. We use the higher resolution DRAO image as a guide to avoid measuring areas contaminated by bright compact sources. We first subtract the average background level, and then measure the mean Stokes $I$, $Q$, and $U$ in small ($28.5'\times9.5'$) regions on the brightest regions of the Northern and Southern filaments. The bias-corrected polarized intensity is found by $P=\sqrt{Q^2+U^2-(1.2\sigma_{QU})^2}$, where $\sigma_{QU}=800~\mu$K. In both filaments, we measure similarly high polarized fractions of $p=40\%\substack{+30 \\ -20}\%$. The high uncertainty is due to the large variability in the background level. The presence of coincident H$\alpha$ filaments indicates that some small amount of the Stokes $I$ flux is from thermal bremsstrahlung radiation. This would decrease the component of Stokes $I$ coming from the synchrotron radiation, making the measured polarized fraction a lower limit. Despite these being a lower limit, this is a very high polarized fraction value indicating the presence of a very compressed and highly coherent magnetic field that is not dominated by turbulence.  

\begin{figure}[!ht]
\centering \includegraphics[width=9cm]{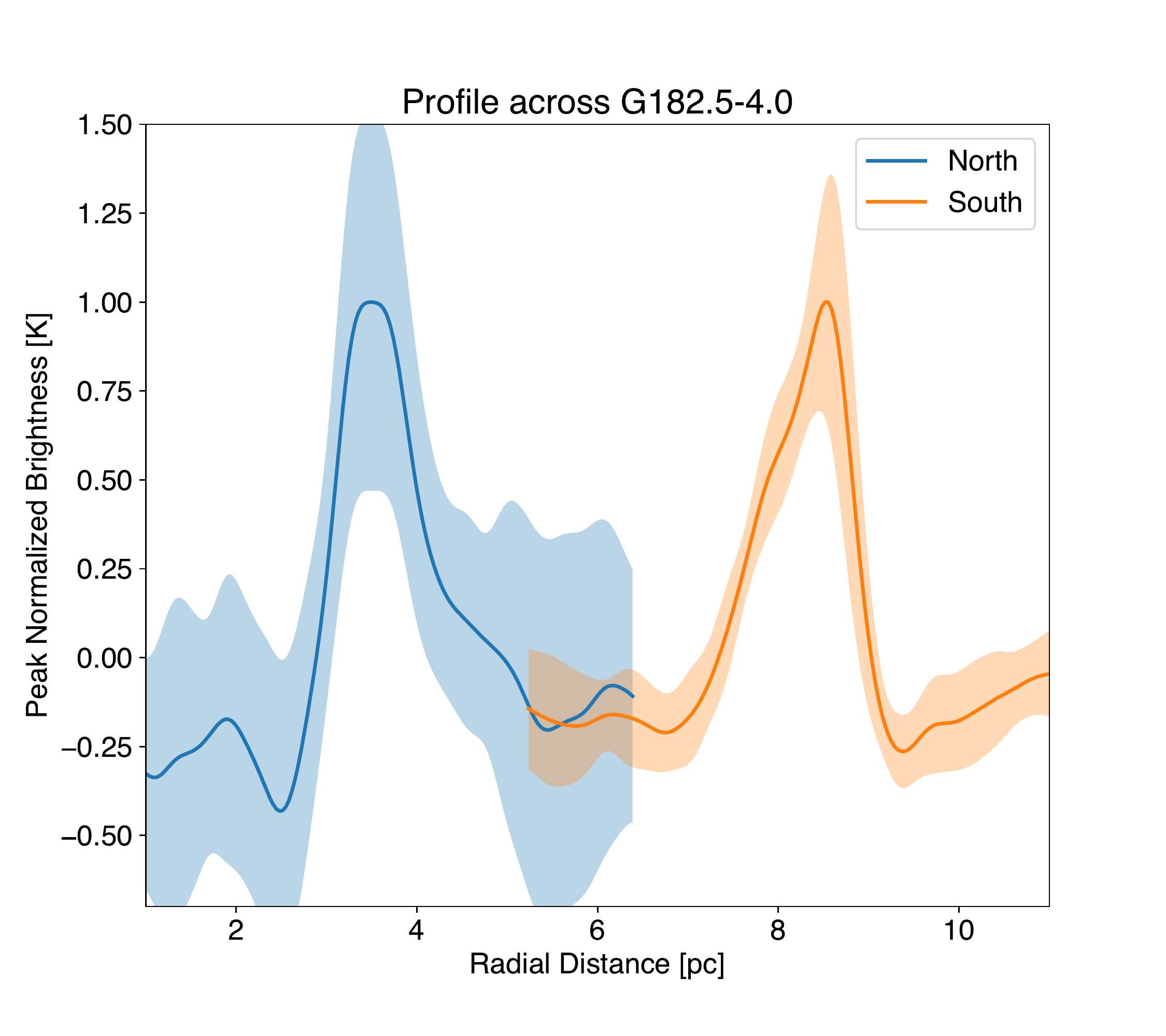}
\centering \includegraphics[width=9cm]{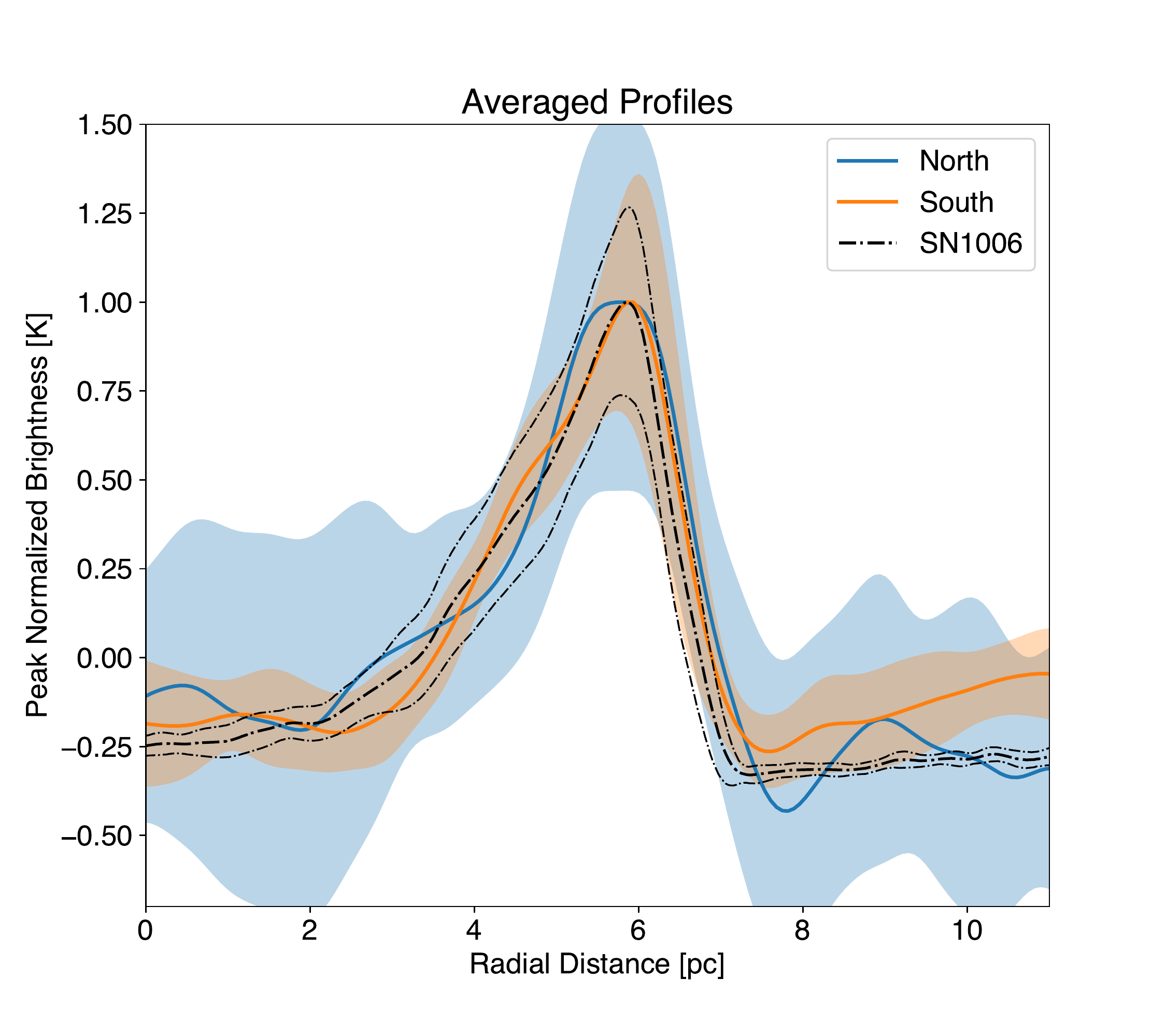}

\begin{scriptsize}\caption{\label{fig:profile}Top: Profile across both the Northern and Southern filaments of G182.5--4.0.  The peak brightness has been normalized to 1. The shading shows the 1$\sigma$ uncertainty on the average. Bottom: Profile across both the Northern and Southern filaments of G182.5--4.0 shown in comparison to the profile of the Eastern limb of SN1006. Here the Northern filament profile has been flipped such that the centre of the shell is at a distance of 0~pc.  The peak brightness has been normalized to 1. As in the top panel, the shading (upper and lower dashed lines in the case of SN1006) shows the 1$\sigma$ uncertainty on the average.
}
\end{scriptsize} 
\end{figure}

\begin{figure}[!ht]
\centering 

\includegraphics[width=8cm]{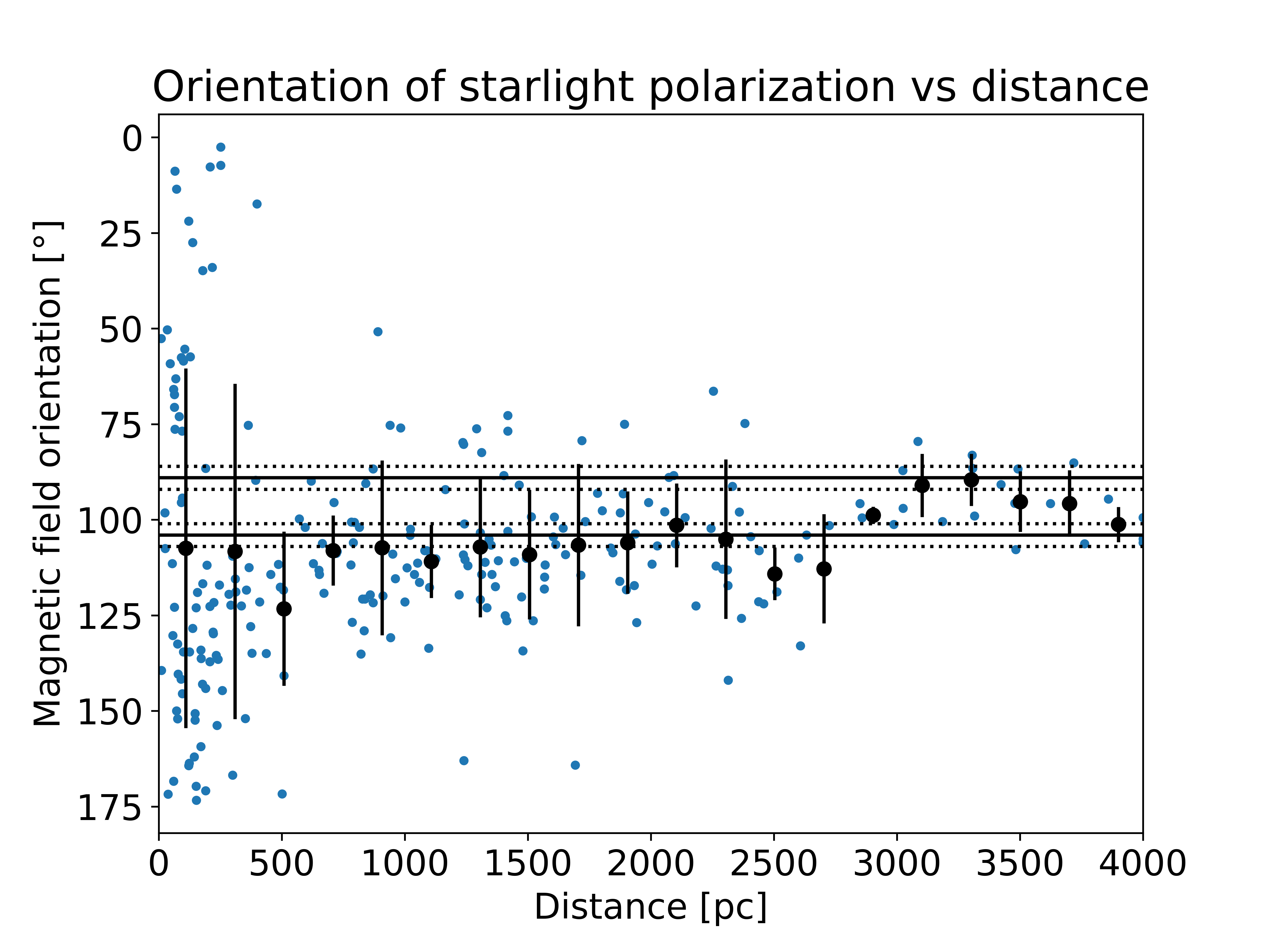}

\begin{scriptsize}\caption{\label{fig:starlight-pol}Plot of the orientation angle vs distance for stars in the catalogue of \citet{2021ChA&A..45..162M}. The horizontal lines represent the orientation of the two parts of G182.5--4.0 with uncertainty (dotted lines). The black points show the mean and standard deviation of the orientation angles using 200~pc bins. 
}
\end{scriptsize} 
\end{figure}

\begin{figure*}[!ht]
\centering \includegraphics[width=5.9cm]{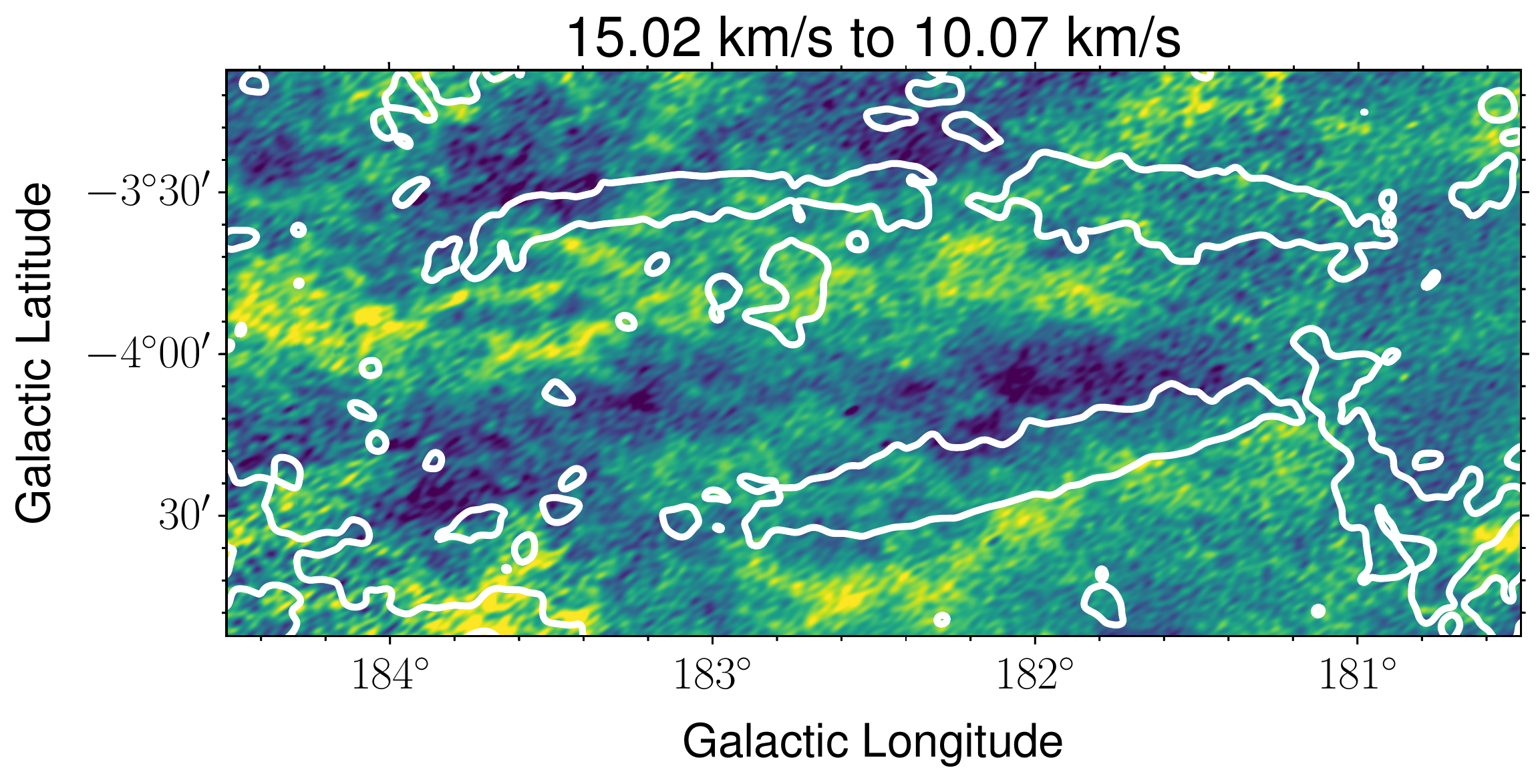}
\centering \includegraphics[width=5.9cm]{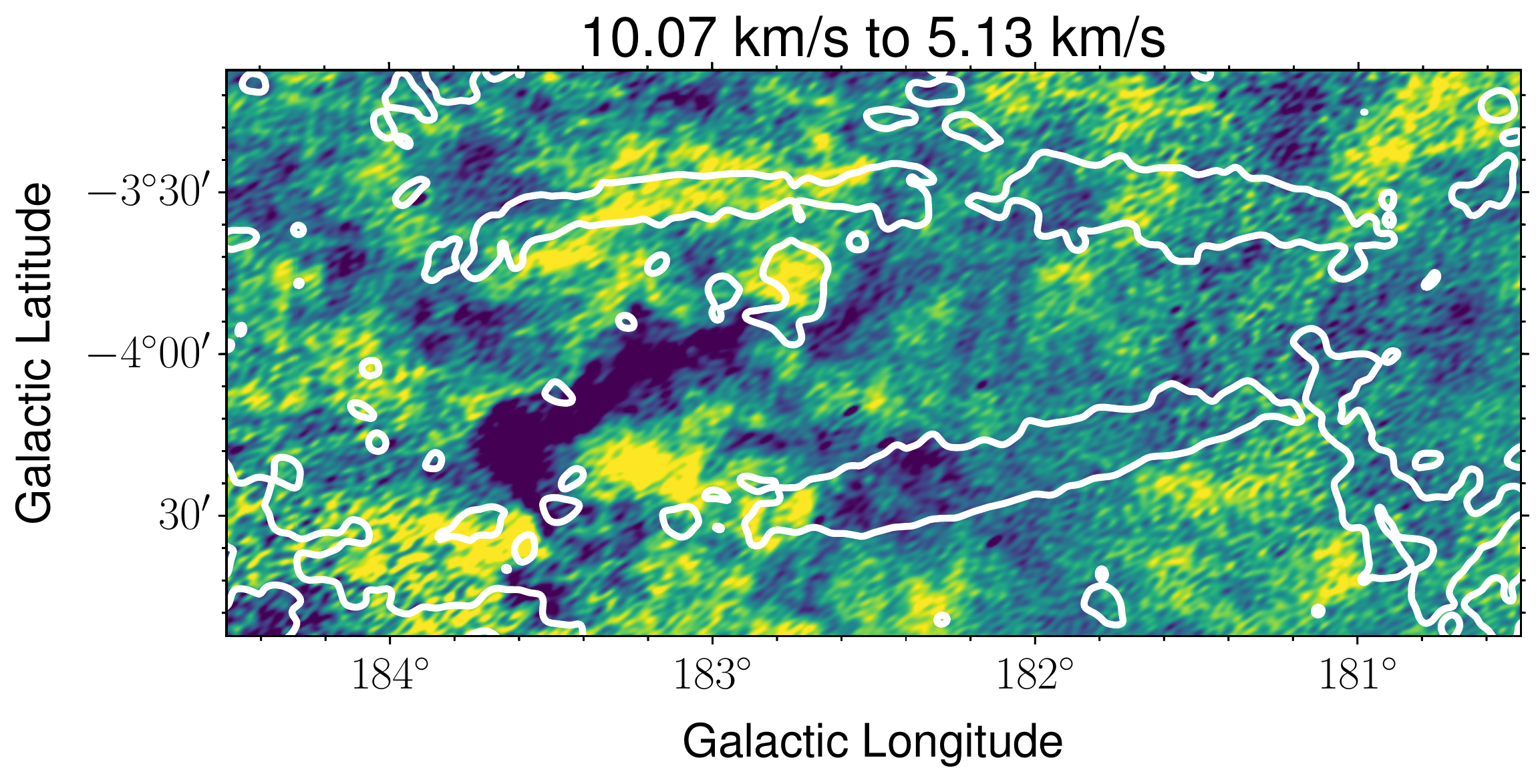}
\centering \includegraphics[width=5.9cm]{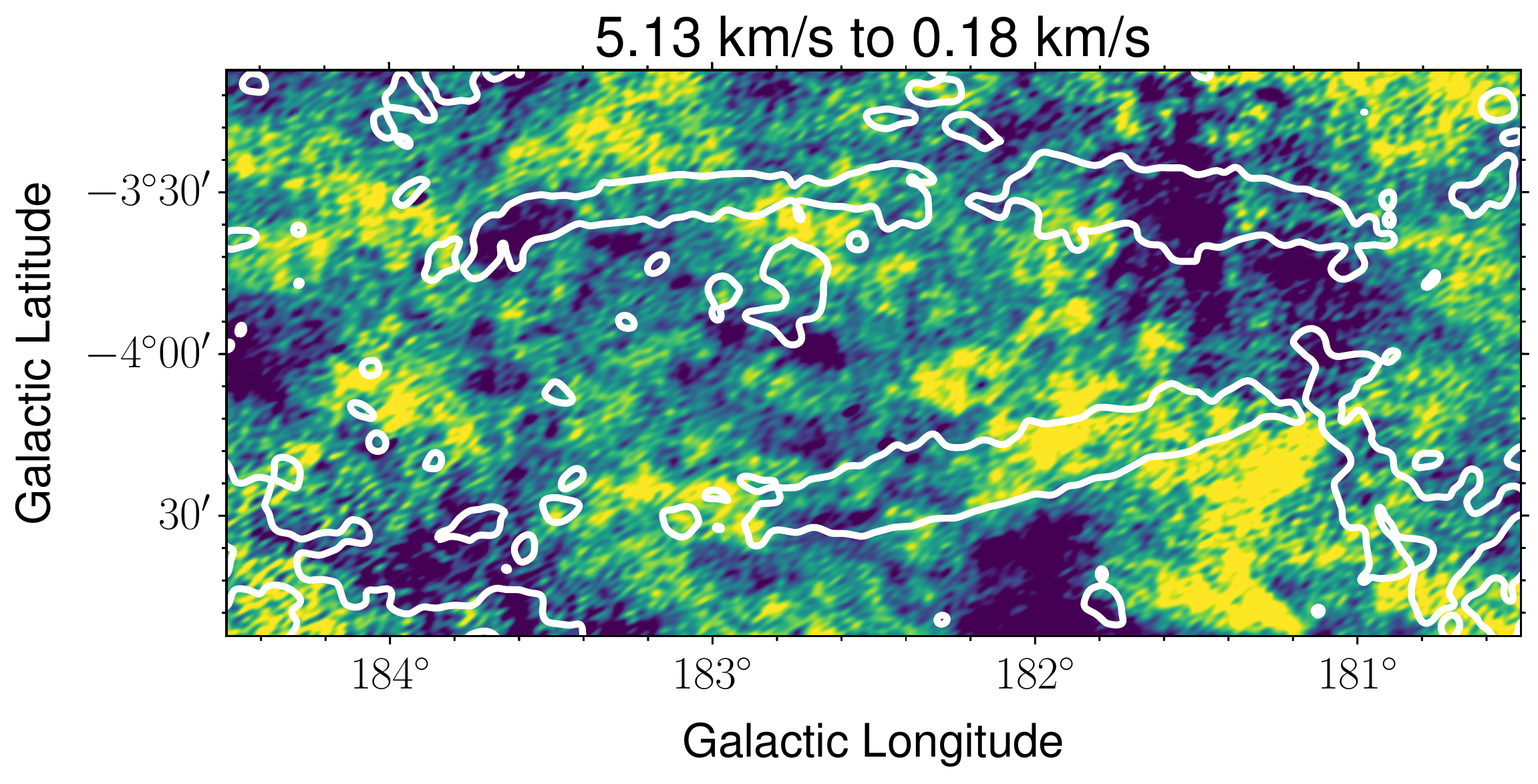}

\centering \includegraphics[width=5.9cm]{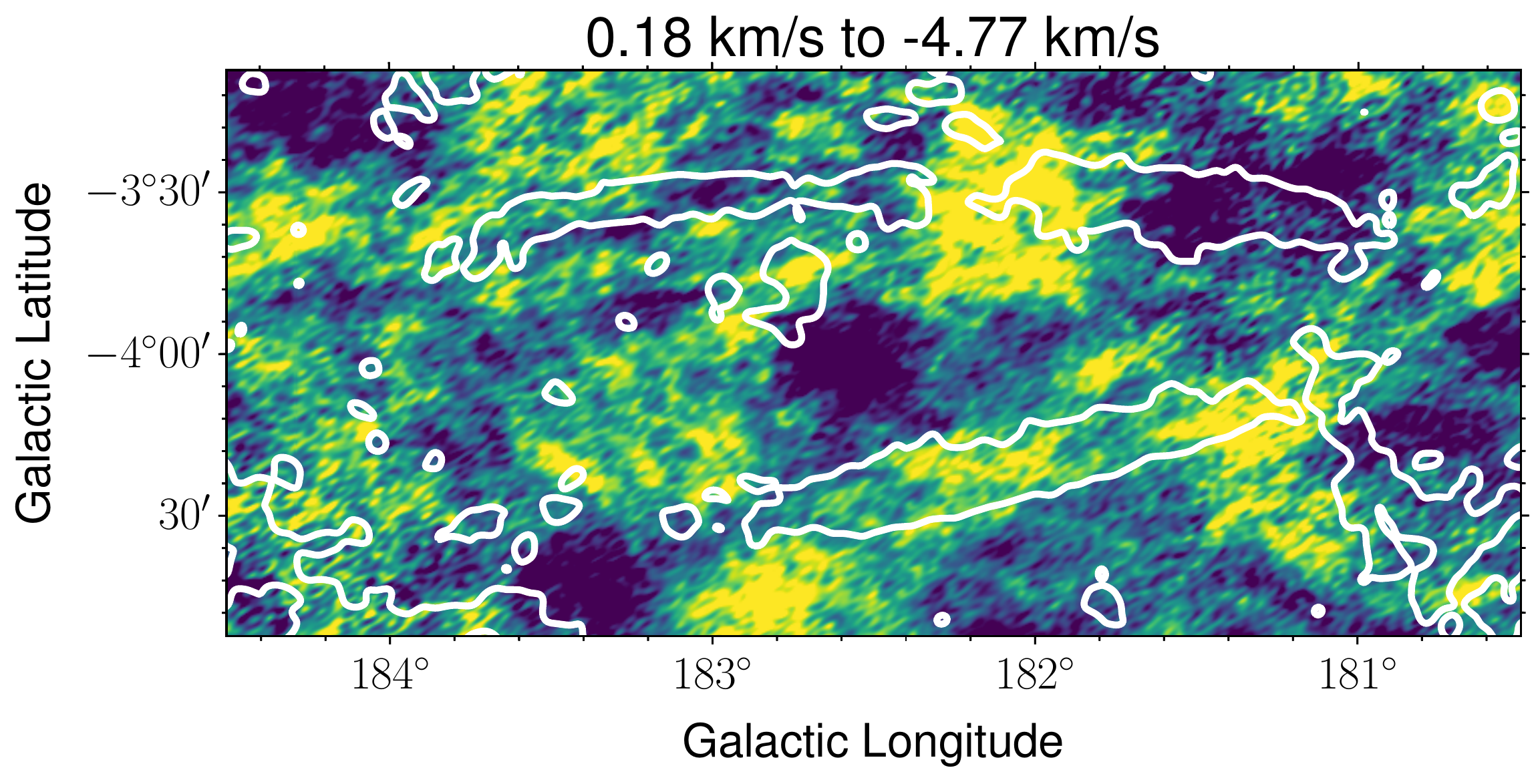}
\centering \includegraphics[width=5.9cm]{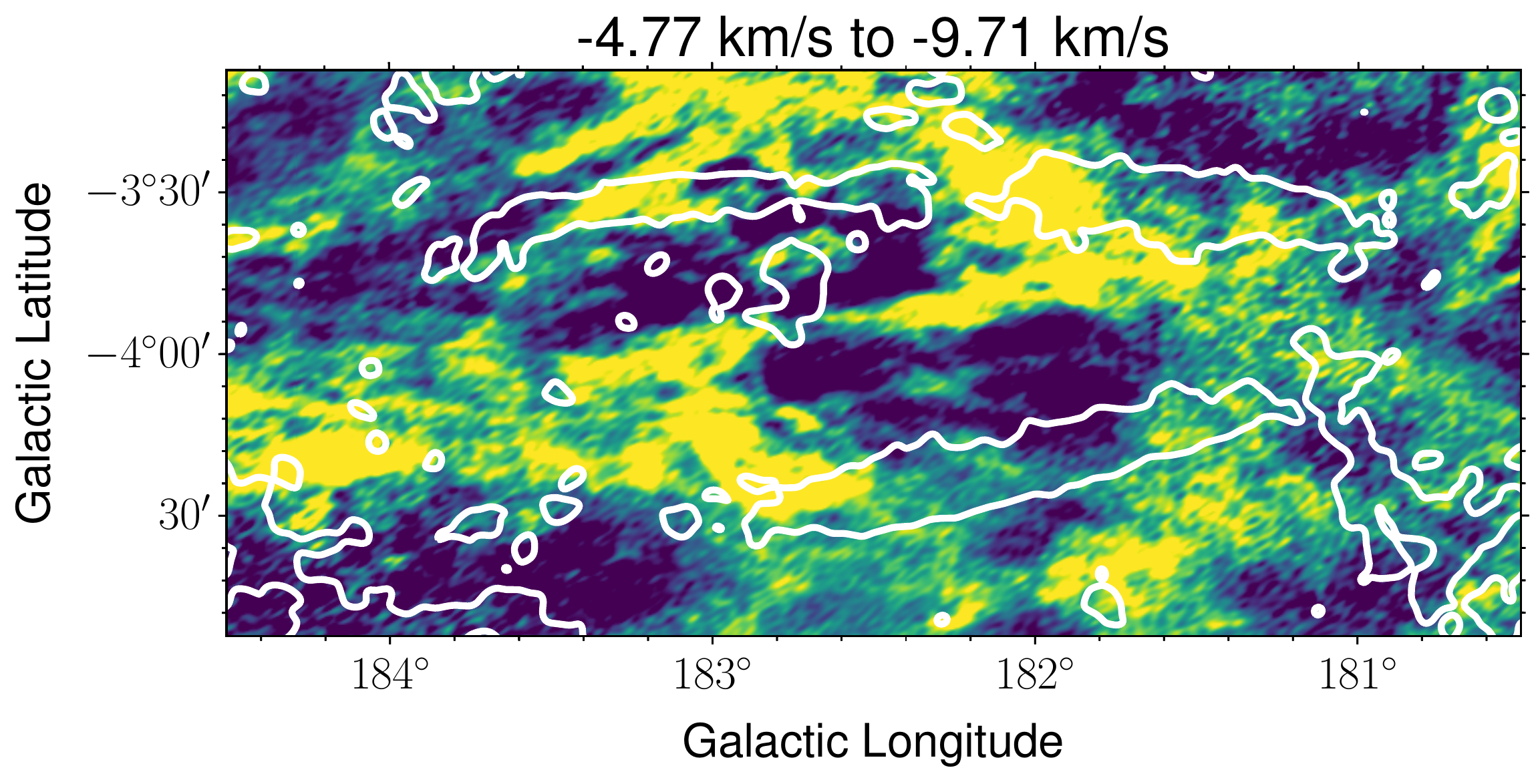}
\centering \includegraphics[width=5.9cm]{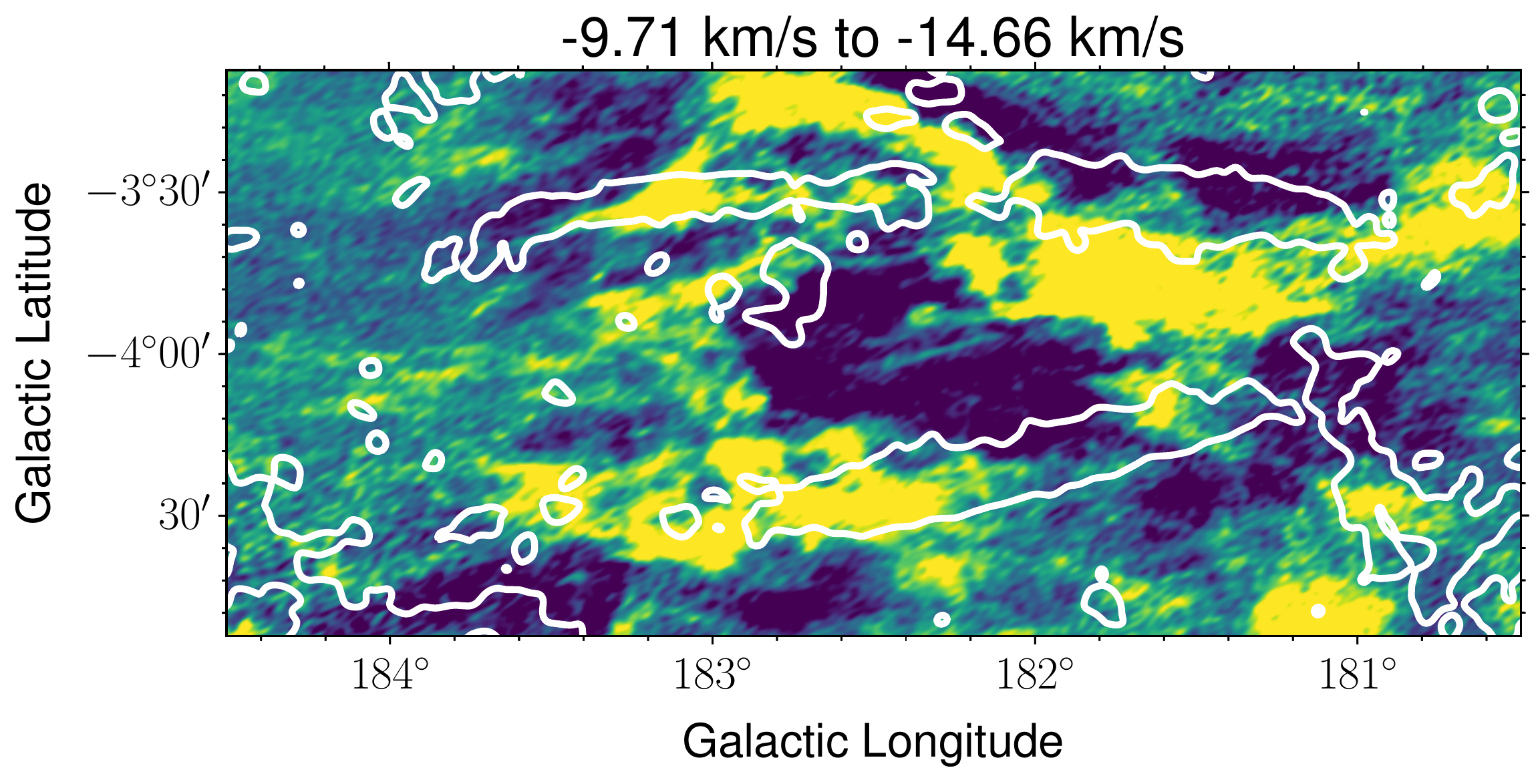}

\centering \includegraphics[width=5.9cm]{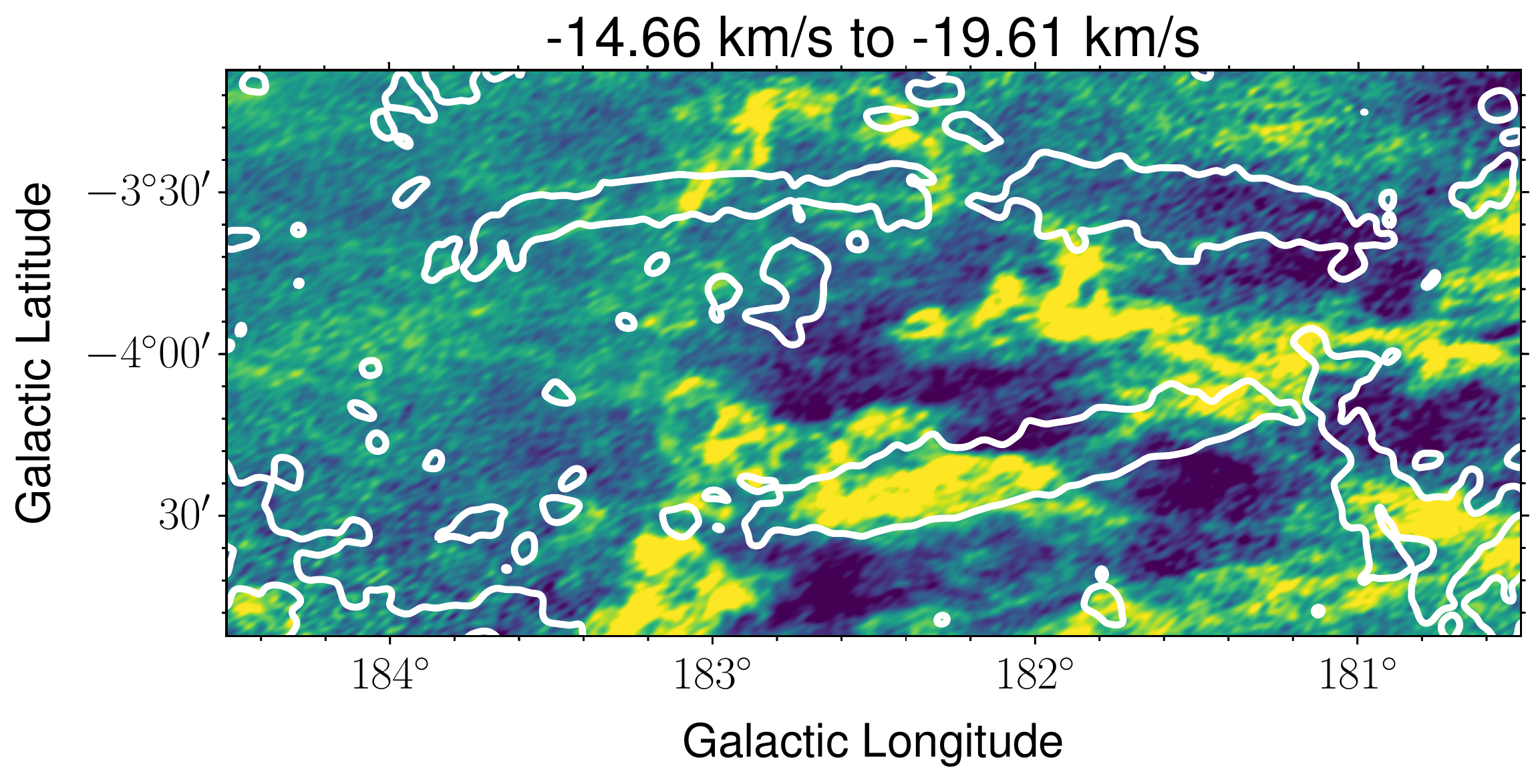}
\centering \includegraphics[width=5.9cm]{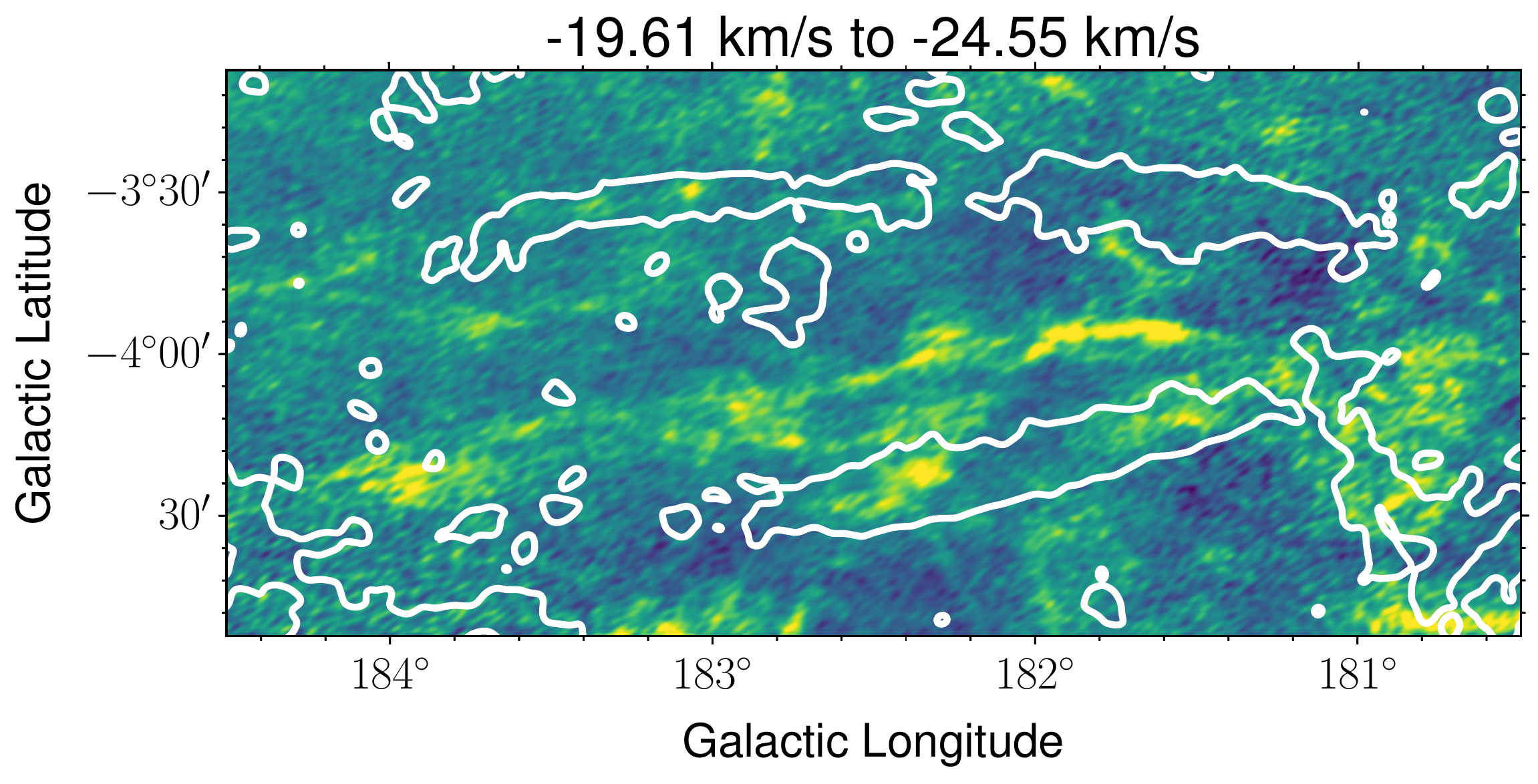}
\centering \includegraphics[width=5.9cm]{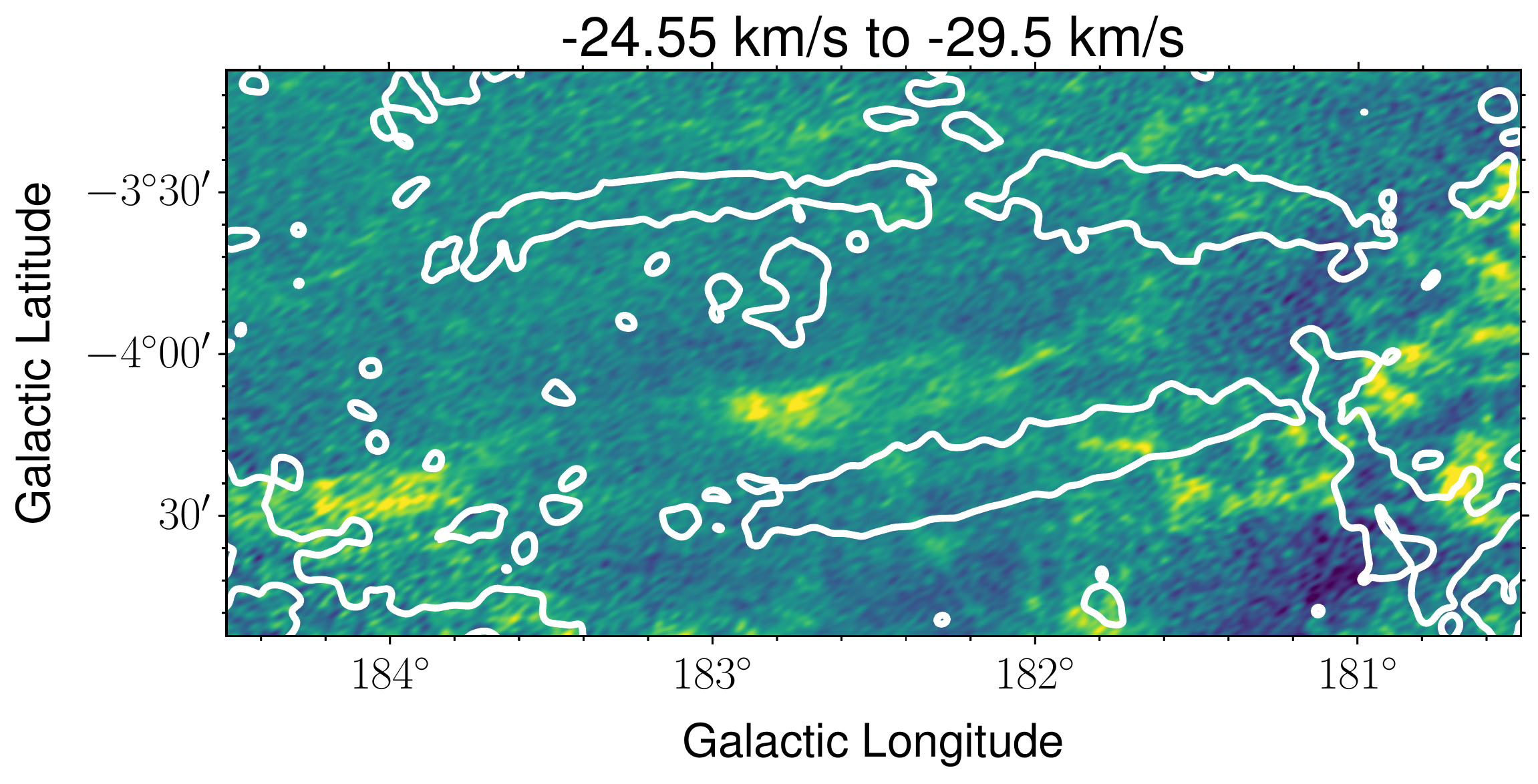}
\begin{scriptsize}\caption{\label{fig:HI-channel}HI velocity maps showing a range of velocities from 15.02~km~s$^{-1}$ to --28.67~km~s$^{-1}$. Each subplot has been averaged over six velocity channels, showing a total range of 4.95~km~s$^{-1}$. Each map is shown with the same linear brightness scaling from --0.12 (dark blue) to 0.12 K (yellow). The white contours show the position of the Stokes $I$ emission of G182.5--4.0.
}
\end{scriptsize} 
\end{figure*}

\begin{figure}[!ht]
\centering 

\includegraphics[width=8cm]{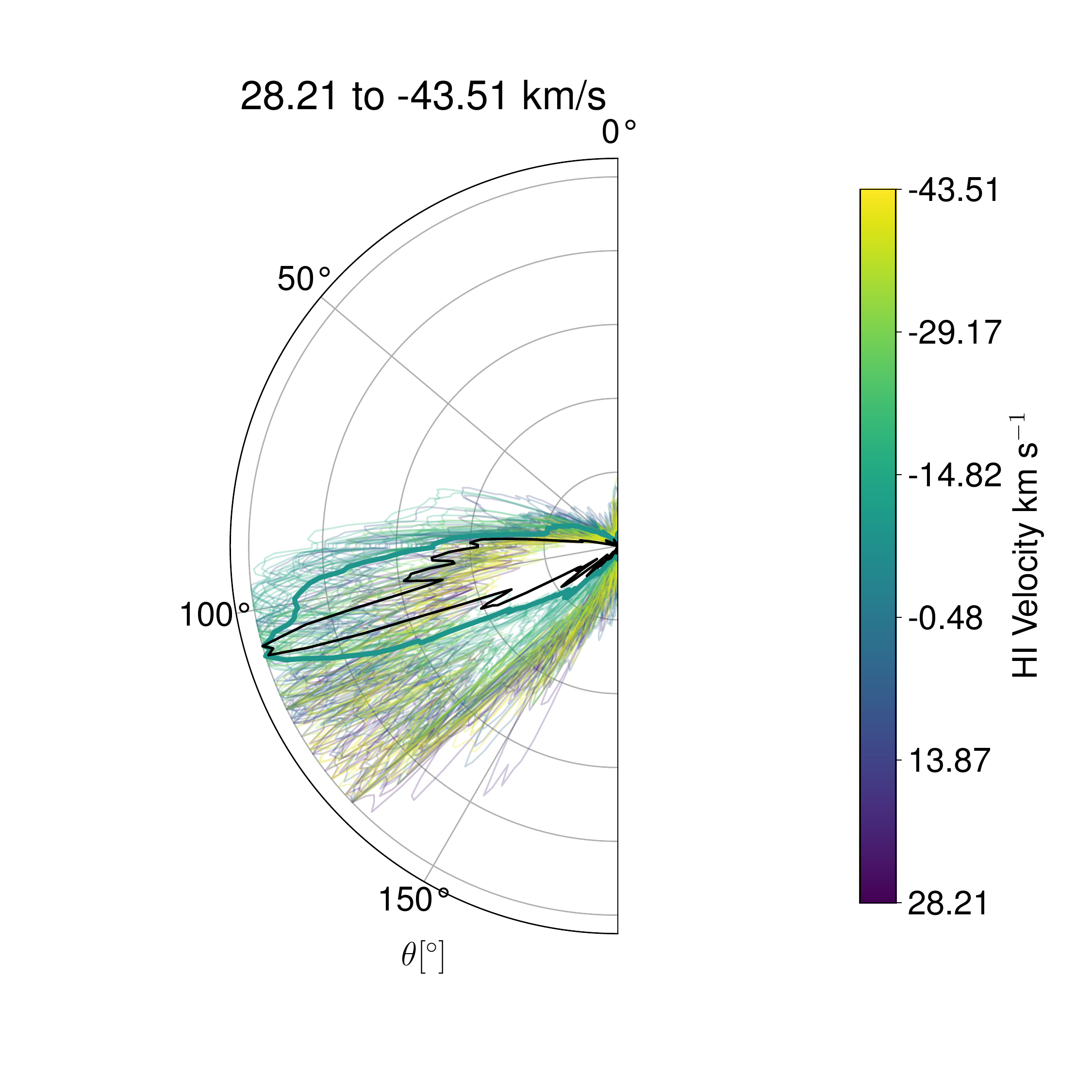}

\begin{scriptsize}\caption{\label{fig:polar-plots}The distribution of RHT orientation angles of HI structures, measured from Galactic north through east, for velocities from 28.21 km~s$^{-1}$ to --43.51 km~s$^{-1}$. Each curve is an integration over all of the pixels enclosed in a region from $180.75^\circ<l<184.00^\circ$ and $-3.33^\circ>b>-4.50^\circ$, and for a particular velocity channel. The black curve shows the result from the Stokes $I$ image. The thicker cyan curve is for an HI velocity channel at --8.8880 km~s$^{-1}$, which shows the greatest agreement with Stokes $I$. 
}
\end{scriptsize} 
\end{figure}

\subsection{\label{sec:profile}Synchrotron Radial profile}

A strong shock, such as that from a SN explosion, forms a compressed shell that can be characterized with a Sedov-Taylor mass density profile \citep{1959sdmm.book.....S}, which gives the observed shell a characteristic brightness profile. This profile has a shallower slope on the interior of the shock and a steeper slope at the outer shock boundary.  Using the DRAO Stokes $I$ data, we make an averaged profile across the Northern and Southern filaments to check if the profile is consistent with the two filaments being on either side of an interior explosion (i.e., the shock travels from a point in the centre outwards) or whether the shock has impacted both filaments from the same direction, as we would expect if the filaments were both located on the same edge of a much larger shell. 

We define two boxes with each filament running through the centre. For the Northern filament, the box is $2.0^\circ\times1.0^\circ$, extending from $181.4<l< 183.4$ and $-4.0<b<-3.0$. For the Southern filament, the image is first rotated by $13^\circ$ so that the box can be defined approximately perpendicular to the filament. This box extends $1.5^\circ\times1.0^\circ$ from approximately $181.2<l< 182.7$ and $-5.0<b<-4.0$. The values are averaged along the long dimension of the box (longitude axis) and then plotted along the short dimension of the box (latitude axis). In the top panel of Fig.~\ref{fig:profile}, it is clear that the profiles of the Northern and the Southern filaments have an opposite shape, indicating that the steeper slope is located at the outer edge of each filament. This is consistent with the origin of the shock being centrally located, between the two filaments. Here the angular dimension is converted to a physical scale using an assumed distance of 400~pc. 

We also compare the profiles to that from a canonical SNR, SN1006. Using archival 843 MHz data from the Sydney University Molonglo Sky Survey (SUMSS)\footnote{data obtained from SkyView, \url{https://skyview.gsfc.nasa.gov}} we make a similar profile of the Eastern limb of SN1006 averaging over a box measuring $0.2^\circ$ in latitude and $0.4^\circ$ in longitude. Assuming a distance to SN1006 of 1.85~kpc \citep[][and references therein]{2019JApA...40...36G}, we find that the physical scale is approximately double that of G182.5--4.0. Thus in order to show the profiles on the same scale, for the bottom panel of Fig.~\ref{fig:profile} we use a distance of 800~pc to convert the angular dimension to a comparable physical scale (instead of 400~pc used in the top panel of Fig.~\ref{fig:profile}). Since the shape of the Sedov-Taylor profile is self-similar, the physical scaling is somewhat arbitrary and does not impact the comparison of the relative shape of the profile. The profiles are nearly identical, supporting the interpretation that the origin of these filaments are from a shock with a centrally located origin such as that from a SN explosion.

\subsection{\label{sec:polstars} Polarized star catalogue}

\citet{2021ChA&A..45..162M} cross-matched the Gaia Data Release 2 (DR2) catalogue with the starlight polarization catalogue from \citet{2000AJ....119..923H} to obtain precise distance measurements together with polarization information for 7613 stars. The polarized orientation of a particular star probes the integrated orientation of the magnetic field along the LOS to the star, weighted by the dust mass. 
In Fig.~\ref{fig:starlight-pol}, we plot the magnetic field orientation vs distance for a region surrounding G182.5--4.0, between $-10^\circ<b<5^\circ$ and $160^\circ<l<200^\circ$

For very nearby stars ($\lesssim300$~pc), there is a large range of orientations, but between 500~pc and 2.5~kpc, the angles are fairly consistent. At relatively large distances ($\gtrsim2.5$~kpc), there is a small, but noticeable shift in the orientation of the starlight polarization. The filament orientations are consistent with a large range of distances,  so it is not possible to constrain the distance with this measurement. However it does suggest that either the dust polarization is dominated by a foreground structure, such as the local bubble wall and/or the wall of the Orion-Eridinaus superbubble, or that there is a remarkably coherent magnetic field in this direction. In both scenarios, given the similar orientation of the filaments with the dust, it suggests the orientation of the filaments are influenced by the ambient Galactic magnetic field. The former scenario supports an association with the Orion-Eridinaus superbubble wall. Although, even though the magnetic field traced by starlight (i.e., dust) polarization is rather uniform along the LOS, it is difficult to make a strong conclusion about the magnetic field traced by the synchrotron emission because synchrotron and dust have different weightings of the field along the LOS.

\begin{figure*}[!ht]
\centering \includegraphics[width=18cm]{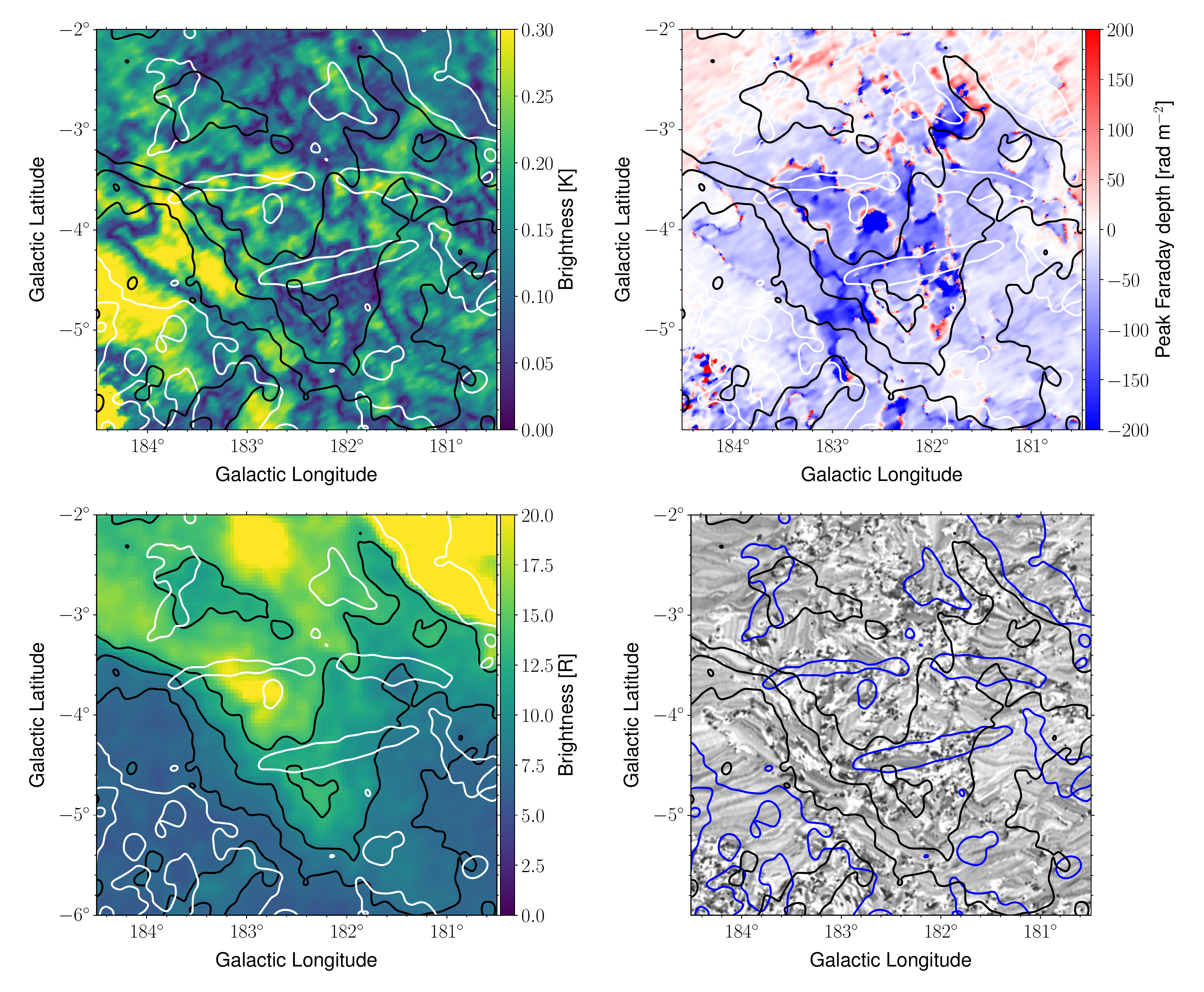}
\caption{\label{fig:RM-halpha}Top left: Map of the peak polarized intensity, $\bf{F}(\phi_\text{peak})$ from GALFACTS. Top right: Map of peak Faraday depth, $\phi_\text{peak}$. Bottom left: WHAM H$\alpha$ map towards the region around G182.5--4.0. Bottom right: LIC drapery lines showing the plane-of-sky magnetic field orientation ($\chi_\text{src}+90^\circ$). All panels have black contours that show the H$\alpha$ emission at levels of 7, 9, and 13~Rayleighs (R). The white contours (blue in the bottom right panel) are from the DRAO Stokes $I$ data with a level of 0.015~K, showing the position and orientation of the G182.5--4.0 filaments. 
}
\end{figure*}

\subsection{\label{sec:rht}Neutral hydrogen}

In Fig.~\ref{fig:HI-channel} we show a range of the HI spectral channels at velocities from $+15.0$2~km~s$^{-1}$ to $-29.5$~km~s$^{-1}$, noting that \citet{2019A&A...631A..52J} find that the velocities of the HI lines associated with the Orion-Eridinaus superbubble range from $+18$ to $-15$~km~s$^{-1}$. Each subplot has been averaged over a velocity range of 4.95~km~s$^{-1}$, and includes Stokes $I$ contours that show the position of the filamentary structure of G182.5--4.0. A visual inspection of these plots indicates that for some velocities there is evidence of elongated HI structures that appear parallel to the Stokes $I$ filamentary emission. That is, there is a similarity in the orientation of the structures but they are not co-located. The most prominent HI filament is located between the synchrotron filaments, as particularly visible in the centre panel of Fig.~\ref{fig:HI-channel} (velocities of $-4.77$~km~s$^{-1}$ to $-9.71$~km~s$^{-1}$) but also to lesser extent at some other velocities (e.g., $-19.61$~km~s$^{-1}$ to $-24.55$~km~s$^{-1}$).

The Rolling Hough Transform (RHT), an algorithm that quantifies the linearity and spatial coherence of structures in images, has been applied to measure the orientation of linear structures in HI data \citep{2014ApJ...789...82C}. We apply the RHT to both the HI velocity cube and DRAO 1.4~GHz Stokes $I$ data. The RHT uses two parameters, a window length (WLEN) and a smoothing radius (SMR). These parameters control the scale (width and length respectively) of the filamentary structures that are quantified by the RHT. We use WLEN$=0.5^\circ$ (101 pixels) and SMR$=0.055^\circ$ (11 pixels). These values approximately correspond to the width and separation of the synchrotron filaments. 

In Fig.~\ref{fig:polar-plots}, we show the average orientations of HI and Stokes $I$ structures over a region from $180.75^\circ<l<184.00^\circ$ and $-3.33^\circ>b>-4.50^\circ$. The range of angles from the Stokes $I$ image, shown in black, has two closely spaced peaks at 105.7$^\circ$ and 107.4$^\circ$, with lesser peaks at 89$^\circ$, 94$^\circ$, and 99$^\circ$. Fig.~\ref{fig:polar-plots} also shows a selection of Galactic HI velocities from 28.21 km~s$^{-1}$ to --43.51 km~s$^{-1}$. The channel at --8.8880 km~s$^{-1}$ has peaks that are exactly identical to those in Stokes $I$, and is also consistent with the by-eye measurements from Sec.~\ref{sec:properties}. Since the orientation of the HI structures are aligned with the synchrotron filaments, it suggests that either they have the same origin, or that they are simply both tracing the ambient Galactic magnetic field.

\subsection{\label{sec:rmsynth}Rotation Measure Synthesis}

We use the RM-Tools package \citep{2020ascl.soft05003P} to perform 3D RM-synthesis on the GALFACTS data. For each pixel, we find the peak Faraday depth, $\phi_\text{peak}$, the peak polarized intensity, $\bf{F}(\phi_\text{peak})$, and the polarization angle at the source (i.e., the derotated polarization angle), $\chi_\text{src}$, which are standard outputs of RM-Tools. The angle is derotated according to Eq.~\ref{eqn:chiobs} using $\text{RM}=\bf{F}(\phi_\text{peak})$ and where $\lambda$ is a reference frequency, which RM-Tools defines as the central wavelength over the band. 
Fig.~\ref{fig:RM-halpha} shows a map of $\bf{F}(\phi_\text{peak})$ (upper left) and $\phi_\text{peak}$ (upper right), and the magnetic field lines ($\chi_\text{src}+90^\circ$) visualized using the line integral convolution (LIC) technique \citep[lower right,][]{10.1145/166117.166151}. 

$\phi_\text{peak}$ is largest in the vicinity of G182.5--4.0, up to a maximum magnitude $\phi_\text{peak}\sim-300$~rad~m$^2$. Elsewhere the magnitude is largely close to zero. $\chi_\text{src}$ is quite variable and we see obvious depolarization in $\bf{F}(\phi_\text{peak})$ (dark regions) where $\phi_\text{peak}$ is high, noted by the rapidly varying pattern in the magnetic field lines. Where $\phi_\text{peak}$ is small, the angles are mostly smooth. 

In Fig.~\ref{fig:RM-halpha} (lower left), we show the diffuse H$\alpha$ emission,  which is proportional to the square of the electron density integrated over the path length (i.e., emission measure), and see that some of its highest values where $\phi_\text{peak}$ is also high, and the contours follow the dark regions in the $\bf{F}(\phi_\text{peak})$ map. Since Faraday rotation is proportional to the integrated product of the LOS magnetic field and electron density, the larger values of $\phi_\text{peak}$ are more likely due to an increased electron density rather than increased LOS magnetic field strength, although both may contribute\footnote{We note that the highest values of the H$\alpha$ emission are in fact located the top-right corner, and the corresponding $\phi_\text{peak}$ values in this location are quite small. This location corresponds to the known SNR called S147, which is located at a further distance of 1.2~kpc \citep{2019JApA...40...36G}. As noted in the text, Faraday rotation is proportional to a complicated integrated product, and thus depending on the magnetic field geometry and path lengths, values of high electron density do not always correspond to high Faraday depths.}. The coherent magnetic field lines (for example, seen in the region around $(l,b)=(184^\circ, -5^\circ)$), have a similar orientation on either side of the H$\alpha$ emission, implying that the magnetic field may also be coherent in the vicinity of G182.5--4.0. These lines are oriented at 105$^\circ\pm5^\circ$, which is similar to the orientation of the G182.5--4.0 filaments. $\phi_\text{peak}<0$ indicates that the net component of the magnetic field is directed away from the observer.

\section{\label{sec:discussion}Discussion}

G182.5--4.0 is located almost directly towards the Galactic anti-centre. Given the Sun's position at $8.15\pm0.15$~kpc from the Galactic centre \citep[][]{2019ApJ...885..131R}, this implies that in this direction, we are looking through a reasonably small path length through to the edge of the Galaxy. Current Galactic magnetic field models \citep[e.g.,][]{Page:2007ce,Sun:2008bw,  Jaffe:wl, Jansson:2012ep} agree that in this direction the mean magnetic field should be oriented both almost entirely parallel to the Galactic plane, and such that it is in the plane of the sky with little component directed towards or away from the Sun.

By the same argument, the rotation of the Galaxy in this direction is expected to be almost entirely perpendicular to the LOS. Thus, any structures seen at particular HI velocities in this direction are expected to be dominated by peculiar velocities, and not due to Galactic rotation \citep{1972A&A....19...51B, 2010ASPC..438...16F}. For the specific location of G182.5--4.0, we also see HI at a broad range of velocities due to the superposition and/or association with the edge of the expanding Orion-Eridanus superbubble \citep[][]{2019A&A...631A..52J}.

Our analysis shows a clear correspondence between the orientation of Stokes $I$ radio filaments, starlight polarization from dust, and HI fibres. The presence of polarized emission indicates the synchrotron filaments must be magnetic in nature. We consider several options to explain their origin.

\subsection{\label{sec:snr}An old SNR?}

G182.5--4.0 could be a SNR given that these are the most common polarized, shell-type objects that can be seen at radio wavelengths. SNRs in the radiative phase are typically optically bright and sources of thermal electrons, which cause depolarization. The old SNR S147, is a typical example. It can be seen in Fig.~\ref{fig:continuum}, located to the north-east of G182.5--4.0, as highly depolarized with a sharp edge. 
The polarized radio emission with coincident H$\alpha$ emission that we observe in G182.5--4.0 supports the old SNR hypothesis. Our analysis of the shape of the radial profile in Sec.~\ref{sec:profile} shows that it has a nearly identical shape to the known SNR, SN1006. This further supports the SNR interpretation.

However, G182.5--4.0 still has a very unusual morphology for an SNR, which are typically round and have a closed geometry (i.e., like two halves of a bubble). In contrast, these filaments are very elongated, and appear to converge only on one end, with only a slight indication that the other end might also curve inwards. Where optical emission is observed in other SNRs, it is usually a complex network of filaments in the shell, rather than the isolated strands seen in G182.5--4.0.  There are some known SNRs that have somewhat similar morphology, like G065.1+00.6 \citep{1990A&A...232..207L} and G296.5+10.0 \citep{1976AuJPh..29..435D}, although these still lack the straight edges and open morphology seen in G182.5--4.0. 

The filaments of G182.5--4.0 appear very thin and straight, and are highly polarized which implies high compression and therefore a highly evolved age. The location of G182.5--4.0 near the edge of the Orion-Eridanus superbubble, as shown in Fig.~\ref{fig:orioneridanus}, suggests a possible association. A superbubble is not created by a single explosion, but rather the combined effects of winds and SNe in different locations within the bubble. Thus, its shape is not expected to be uniformly spherical, and the shaded band in Fig.~\ref{fig:orioneridanus} is only included to show the approximate boundary of the bubble. Thus even though the filaments are not parallel to the edge of the superbubble, we suggest that the magnetic field in this region may experience increased compression in a direction towards the Galactic plane (i.e., near the top of Fig.~\ref{fig:orioneridanus}). If the explosion occurred in a magnetic field with enhanced compression, it may explain the unusual appearance. An association would imply a distance of about 400~pc, and size of $\sim20$~pc, which is somewhat small for an old SNR. For a single SNR we can assume an upper limit for a physical diameter of about 100~pc, which gives an upper limit on the distance of 2 kpc.

The spectral index would be useful to confirm this scenario since we expect SNRs to have a spectral index of $\alpha=-0.5\pm0.2$ \citep{2012SSRv..166..231R}. However, the 160 MHz bandwidth of the GALFACTS data is insufficient for a reliable in-band measurement. We also attempted to measure a spectral index from archival radio data, including 70--231 MHz data from the GaLactic and Extragalactic Allsky MWA Survey \citep[GLEAM,][]{2015PASA...32...25W}, 74 MHz Very Large Array Low-frequency Sky Survey \citep[VLSS,][]{2007AJ....134.1245C}, GMRT 150 MHz All-sky Radio Survey: First Alternative Data Release \citep[TGSS ADR1,][]{2017A&A...598A..78I}, 2.7~GHz data from the Effelsberg telescope \citep{1984A&AS...58..197R} and 4.8~GHz data from the Urumqi telescope \citep{2007A&A...463..993S}. We were unable to get an estimate or even any useful constraints on the spectral index given the faintness of these filaments, artifacts due to the Crab Nebula, and contamination from compact objects in the low resolution data.  Future broader band data with higher resolution are needed to make a spectral index measurement.

\subsection{\label{sec:bowshock}Bow shock?}

Given the suggestive shape, we explore the possibility that G182.5--4.0 is a bow shock nebula. Bow shocks have been observed around neutron stars, in addition to being seen around other massive stars \citep{2021ApJ...922..233O}. Examples of such objects include the guitar nebula, which is associated with PSR B2224+65 \citep{2021ApJ...922..233O} and G359.23--0.82, associated with PSR J1747--2958 \citep{2004ApJ...616..383G}. In these cases, the bow shock is revealed by high-resolution, multi-wavelength observations (including X-rays, radio, and H$\alpha$) in the immediate vicinity of the pulsar/the pulsar wind nebulae (PWN). These objects all have much smaller angular scale than G182.5--4.0, with lengths on the order of a few arcminutes as opposed to several degrees, and the observations show the tip of the bow shock along with a central tail. 
In the case of the guitar nebula, a more extended H$\alpha$ nebula is also revealed. Models of PWN bow shocks \citep[e.g.][]{2004ApJ...616..383G, 2020JPhCS1697a2002P} show a central PWN with a tail surrounded by a region of shocked ISM that fans out with a larger opening angle. To our knowledge an example of such an extended bow shock nebula, without an associated pulsar, has never been found.

\begin{figure}[!ht]
\centering \includegraphics[width=8.5cm]{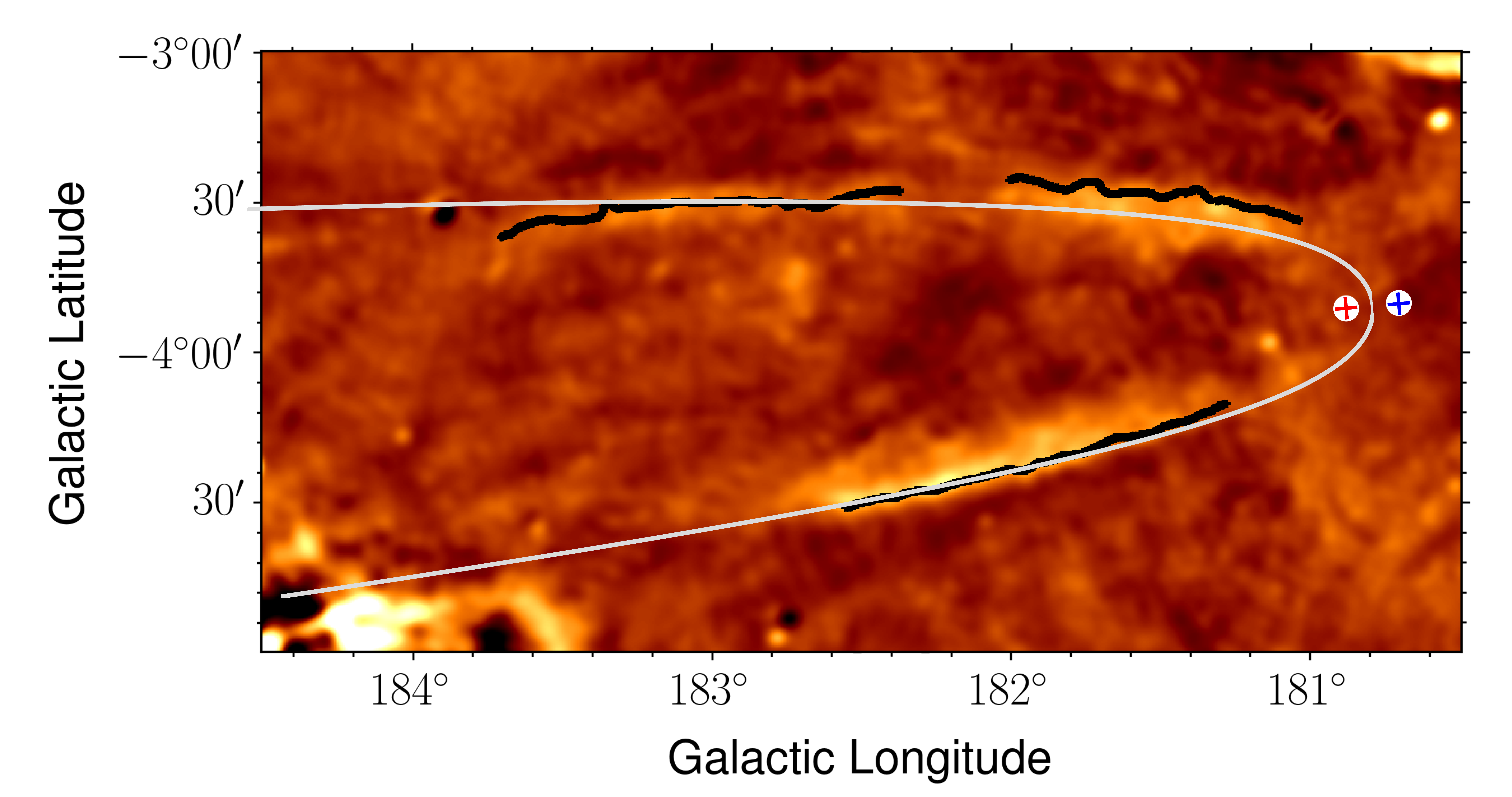}
\centering \includegraphics[width=8.5cm]{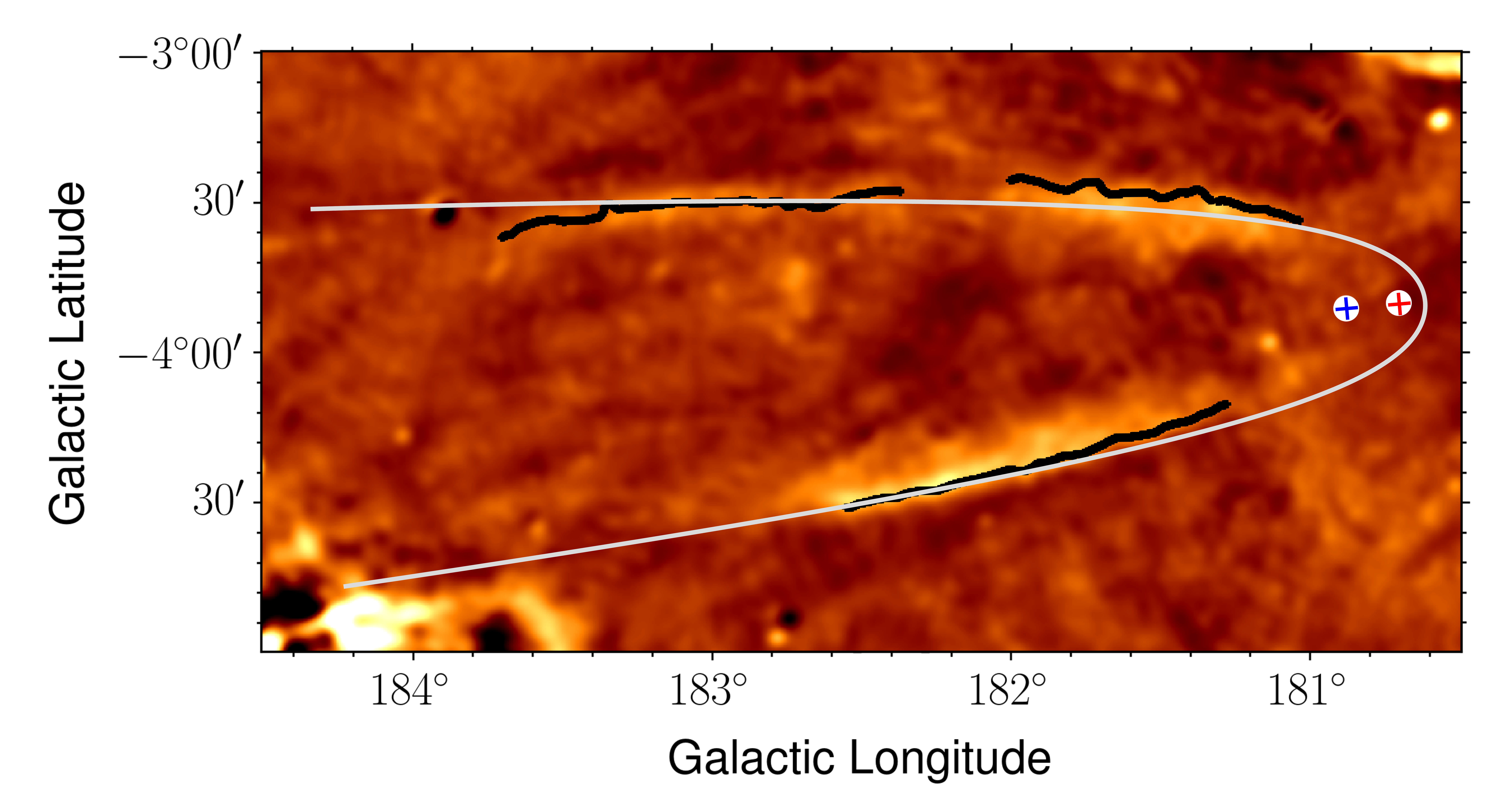}
\centering \includegraphics[width=8.5cm]{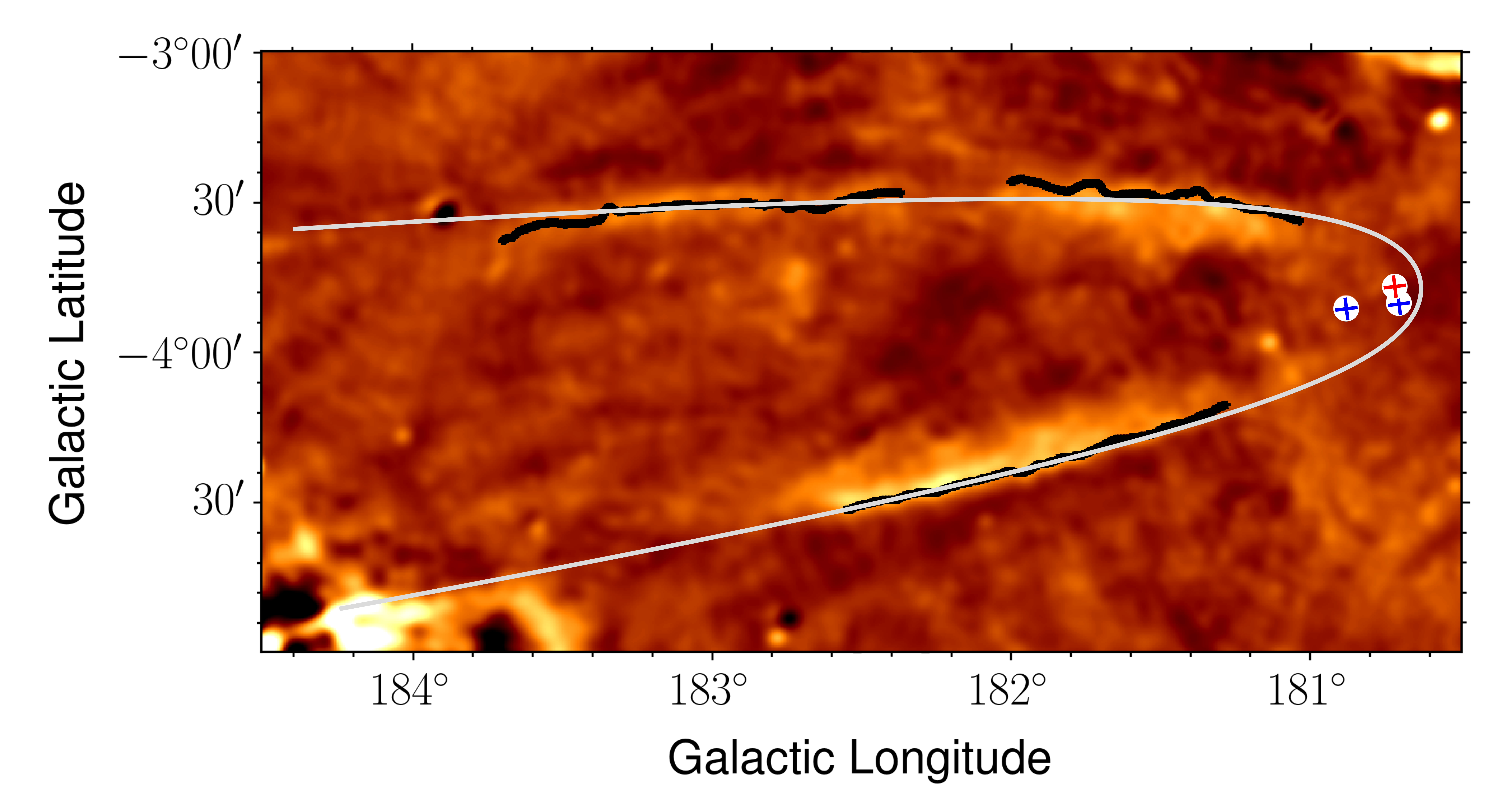}
\begin{scriptsize}\caption{\label{fig:bowshock}Top panel:  Best fit bow shock model (grey line), with the origin (pulsar position) fixed at the location of the X-ray source, 1RXS J053300.7+260827 (shown with a red cross). The black lines are a series of points defined by a data contour with a level of 0.01~K. The fitted value of $R_0=321''$ and the fitted rotation of the model is 5$^\circ$ counter-clockwise. The blue cross shows the position of 1RXS J053239.0+261747. The model is shown overlaid on the DRAO polarized intensity image. Middle panel: Same as the top panel, but for the origin (pulsar position) fixed at the location of the X-ray source, 1RXS J053239.0+261747 (shown with a red cross).  The fitted value of $R_0=310''$ and the fitted rotation of the model is 5$^\circ$ counter-clockwise. Bottom panel: Same as the top panel, but without fixing the pulsar position. The best fit position (shown with a red cross) has RA=5:32:54.5, Dec=+26:18:55, with $R_0=313''$ and the fitted rotation of 7$^\circ$ counter-clockwise. 
}
\end{scriptsize} 
\end{figure}

\begin{figure}[!ht]
\centering \includegraphics[width=8.5cm]{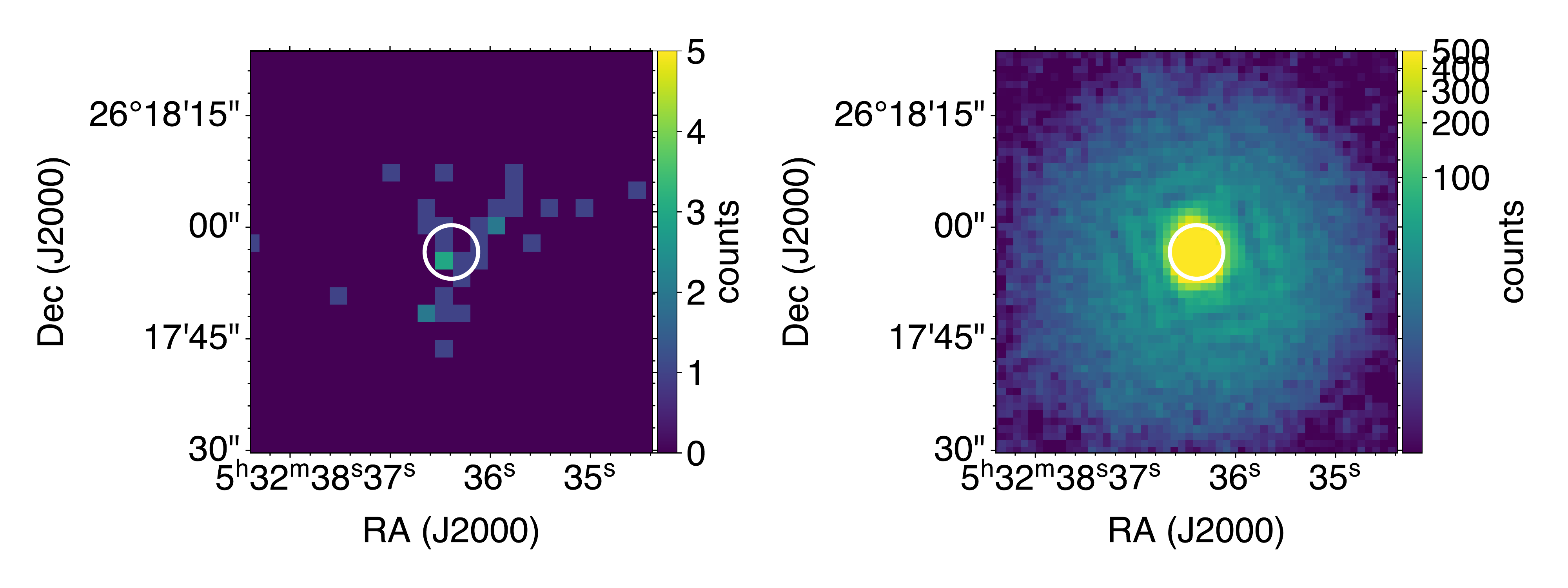}
\centering \includegraphics[width=8.5cm]{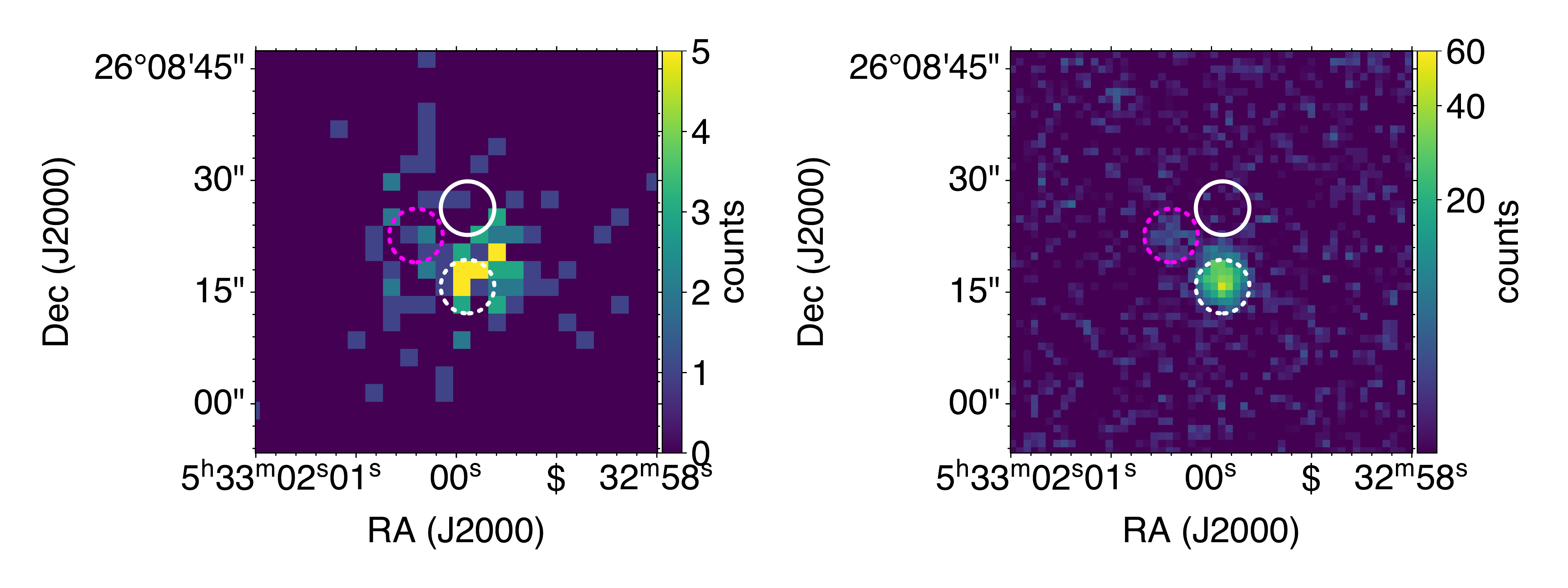}

\begin{scriptsize}\caption{\label{fig:swift}\textit{Swift} spacecraft images of 1RXS J053239.0+261747 (top) and  1RXS J053300.7+260827 (bottom) including X-ray XRT image (left) and ultraviolet UVOT images (right). The white circle in all panels indicates the J2000 catalogue coordinates for the stars HD 244599 (top) and G 98-9 (bottom). The dotted circles in the bottom panel are drawn centred on the sources seen in the \textit{Swift} UVOT image. The white dotted circle in the bottom panel is the presumed current location of the high proper motion star G 98-9. There is a second, unidentified UV component in the bottom right panel (magenta circle) that is offset by $\sim$10$''$ from the X-ray peak.
}
\end{scriptsize} 
\end{figure}

We search for X-ray sources near the intersection point of the two filaments (i.e., what would be the tip of the bow shock) and we find two unidentified ROSAT X-ray sources: 1RXS J053239.0+261747 and 1RXS J053300.7+260827 (see Fig.~\ref{fig:bowshock}). These are found in the ROSAT All Sky Survey with a shallow exposure and only a few counts, making it difficult to draw any conclusions from these sources. 

As follow-up, we obtained a 7255~s exposure, observed on September 7, 2022, with the \textit{Swift} spacecraft (Target ID  15319), including imaging with the X-ray Telescope (XRT) and the Ultraviolet/Optical Telescope (UVOT). While these data, shown in Fig.~\ref{fig:swift}, still have only tens of counts, making detailed analysis challenging, we are able to use them to provide much better localization of the ROSAT sources and thus we can determine if there are any known stellar counterparts. For 1RXS J053239.0+261747, HD 244599, a G0 type star \citep{1993yCat.3135....0C} that is 115.2$\pm$0.2~pc away \citep{2020yCat.1350....0G} is a likely counterpart. For 1RXS J053300.7+260827, there is an M3 type \citep{2004AJ....128..463R} star, G 98-9, with a distance of 41.90$\pm$0.06~pc \citep{2020yCat.1350....0G}. In this case, the catalogued position is well offset from the peak of the X-ray emission. However from examining the UV image, it is clear that the current position of this high proper motion star is coincident with the X-ray emission. For this source, there is a second, unidentified UV component (magenta circle in the bottom panel of Fig.~\ref{fig:swift}) that is offset by ~10$''$ from the X-ray peak. It is highly unlikely that the bow shock could be produced by a M or G-type star, however, there remains the possibility that an unseen binary companion is present, perhaps a massive star that produced the bow shock before exploding as a supernova. However, it is more likely that the bow shock is unrelated to these sources. Further X-ray follow-up, including a spectral analysis is needed to asses whether the observed X-ray emission is consistent with emission from the stellar counterpart. 

In the case of a bow shock, a star would need to traverse the $3^\circ$ length of the object since the SN explosion occurred. Assuming a maximum lifetime of 100~000~years, the minimum speed for such a star is 108~mas/yr (about 210 km/s at 400~pc distance). We searched Gaia DR2 data \citep{2018A&A...616A...1G} but we were unable to locate a star with a fast enough velocity that is also moving in an appropriate direction.

 We fit our data with the analytic form of a bow shock \citep{1996ApJ...459L..31W}

\begin{equation}
    R(\phi)=R_{0}\sqrt{3(1-\phi/\tan\phi)}/\sin\phi
\end{equation}

to find  the ``standoff'' distance, $R_0$. This gives the opening angle of the shell and the position of the origin, which is the location of the pulsar, and also sets the length scale for the shell. $R(\phi)$ is the distance from the pulsar to the edge of the bow shock and $\phi$ is the angle between the origin (e.g., pulsar) and the corresponding point on the shock. In order to fit our data, we define a set of points using a contour at a level of 0.01 K, which traces the outer perimeter of the shock (shown as black points in Fig.~\ref{fig:bowshock}). We fit these points for the parameter $R_0$, in angular distance units, using $1''$ increments and a least-squares fitting approach. We perform the fits over the range $-170^\circ<\phi<170^\circ$ in two ways: first by fixing the pulsar position (origin) at the position of each of the ROSAT X-ray sources, and second by allowing the position of the origin to freely vary (in 1 pixel increments, where 1 pixel = 18$''$). We also allow the image to freely rotate in $1^\circ$ increments to find the best-fit rotation angle, $\theta$. 

The best fits are shown in Fig.~\ref{fig:bowshock}, with an overall best-fit pulsar position of RA=5:32:54.5~s, Dec=+26:18:55, with $R_0=313\pm1''$ and $\theta=7\pm0.5^\circ$. The X-ray source 1RXS J053239.0+261747 is about $3'.6$ from the best fit position. 

If we assume a distance equal to that of the Orion-Eridanus superbubble ($\approx400$~pc), then  $R_0=0.6$~pc. This value of $R_0$ is large compared to many known bow shocks \citep[][]{2004ApJ...616..383G}, but it is in excellent agreement with the value found for the Vela X1 bow shock where $R_0=0.57$~pc \citep{2018MNRAS.474.4421G}. In that case the distance is around 2~kpc, making the angular size five times smaller, and the nebula is only detected much closer to the pulsar, within $-90^\circ\lesssim\phi\lesssim90^\circ$. Using the MeerKAT telescope, \citet{2022MNRAS.510..515V} report 1.3 GHz radio emission from the head of this bow shock, which also has an associated H$\alpha$ component. Radio emission have only been observed around a few bow shock nebulae. At this distance the length of the shell is $\approx19$~pc, which is quite long, but it is about the same distance as another known bow shock nebula, the Mouse, which has a length of $\approx17$~pc  \citep[assuming a distance of 5~kpc, ][]{2004ApJ...616..383G}.

The lack of central emission in G182.5--4.0, and the possibility of a large physical size suggests that if it is a bow shock, it is in a different, most likely later, evolutionary stage from the other examples of this type. The very strong evidence connecting the filaments of G182.5--4.0 to the surrounding Galactic ISM implies that its morphology has been shaped by the ambient Galactic magnetic field. If the bow shock scenario is correct, this would be the first time that such a connection has been shown. 

There is also the possibility of a hybrid scenario, where both an old SNR and a bow shock nebula are at play. In this case, the radio emission would arise from an old SNR rejuvinated by the relativistic wind of a fast-moving pulsar at the stage of leaving its SNR shell. Such a scenario has been observed from the old SNR CTB~80 characterized by long radio ridges intersecting at the pulsar's location where a compact bow shock nebula has been also detected  \citep{1995ApJ...439..722S, 2003AJ....126.2114C}.

As discussed in the previous section, spectral index is useful diagnostic to distinguish a bow shock from the SNR scenario since a bow shock is expected to have a relatively flat spectral index of $-0.3\lesssim\alpha\lesssim0$ \citep{2006ARA&A..44...17G}.  Follow-up observations are needed to measure the spectral index, however, the spectral index measurements to which we are referring have been made in a different part of the bow shock, and it is not clear that this flat spectral index would apply to this extended, shocked ISM structure that we observe. Follow-up X-ray observations at the location of the possible pulsar are crucial to determine their nature and either confirm or rule out the bow shock scenario.

\subsection{\label{sec:filaments}A relic fragment related to multiple SN explosions?}

 The Orion-Eridanus superbubble is located below the Galactic plane at a distance of $\sim400$~pc, and with a diameter of $\sim300$~pc, is thought to have been created by winds or SNe from the Orion OB association \citep{2015ApJ...808..111O}. Its expansion towards the Galactic plane may be providing additional compression to the Galactic magnetic field in this region. This may be a factor that is influencing the morphology of the filaments in the two scenarios (SNR or bow shock) already discussed.  But a third scenario, that G182.5--4.0 is a fragment in the ISM left over from repeated compression and excitation from successive SN explosions within the superbubble is also a possibility. If the relic fragment scenario is true, we should expect to find more isolated, non-thermal radio filaments as radio surveys improve in sensitivity and resolution.

\section{\label{sec:conclusions}Conclusions}

We have presented the properties of the newly discovered radio filamentary structure that we call G182.5--4.0. We have considered several possible scenarios that could explain this structure including an old SNR, a bow shock nebula from a neutron star, or a relic fragment left over from one or more very old SN explosions. Given that there is evidence supporting both the old SNR and a bow shock scenarios, we consider that G182.5--4.0 is likely a hybrid-type object where we are seeing an old SNR that may be lit up by the wind of a fast pulsar, and that this is a relatively nearby object located in an unusual environment that has been highly compressed by Orion-Eridanus superbubble. 

We find the filaments are oriented at 89$^\circ\pm3^\circ$ (Northern filament), and 104$^\circ\pm3^\circ$ (Southern filament). Starlight polarization shows consistent orientations for a wide range of distances. Using the RHT technique, we show that HI at a velocity of --8.88~km~s$^{-1}$ has an orientation consistent with the Stokes $I$ emission.

We use 3D RM-Synthesis on the GALFACTS data to measure $\bf{F}(\phi_\text{peak})$, $\phi_\text{peak}$, and $\chi_\text{src}$ near G182.5--4.0. The Faraday rotation appears to have a peak magnitude in close vicinity to G182.5--4.0. Through comparison with H$\alpha$, we conclude that these high negative Faraday depth values are most likely due to a foreground thermal electron cloud.

We make the following conclusions:

\begin{enumerate}
  \item G182.5--4.0 is polarized, with a high linear polarized fraction of $40\substack{+30 \\ -20}\%$. This indicates that the origin of the emission is synchrotron radiation, which implies the presence of relativistic cosmic ray electrons in a magnetic field.
  \item The radial profile of the filaments are consistent with a supernova explosion that was centrally located between the two filaments.
  \item From its straight morphology and high polarized fraction, the magnetic field of G182.5--4.0 must be highly compressed and not dominated by turbulence.
  \item Given that we consistently find $\phi_\text{peak}<0$ over a large spatial area, on average, the magnetic field must be directed away from the observer towards this LOS.
  
  \item The filaments are coincident with an edge of the Orion-Eridanus superbubble, which could be responsible for the high compression of the Galactic magnetic field towards the Galactic plane in this region. An association would imply a distance of $\sim$400~pc.
  \item The shape of the filaments are well fit by a typical bow shock shape, with $R_0=313\pm1''$. At a distance of 400~pc, $R_0=0.6$~pc.
  
  \item We have established that there is a strong connection between the orientation of G182.5--4.0 and the Galactic magnetic field direction in the surrounding region. This is supported by starlight polarization and neutral hydrogen. In the case of an association with the Orion-Eridanus superbubble, a connection is not surprising as we would expect the compression in the bubble wall to effect all of the magnetic field tracers in a similar manner. However, such a connection is unexpected in the case of a bow shock nebula since we expect the shape to be dominated by the velocity of the originating object (i.e., star or pulsar). Thus if G182.5--4.0 is a bow shock, then either the magnetic field alignment is purely coincidental, or there is a link between the kinematics and the magnetic field orientation. 

\end{enumerate}

Targeted observations from the Effelsberg telescope or the VLA, would allow for more detailed analysis of the morphology and an accurate spectral index measurement. Spectral index is an important clue to provide additional evidence to confirm the true origin of this object. Further follow-up X-ray observations of the compact sources near the tip of G182.5--4.0 are necessary to confirm or exclude the bow shock scenario. 

These filaments may represent some new class of objects, either as some hybrid between a SNR and a bow shock nebula and/or the result of repeated compression in the ISM due to successive SN explosions. Filaments such as these may be commonplace in the Galaxy. Studying and locating distances to such filaments could become an important tool to help trace the Galactic magnetic field and construct a true 3D picture of it. Future higher resolution and wide bandwidth radio observations such as those from the LOFAR Two Meter Sky Survey \citep[LOTSS, ][]{2019A&A...622A...1S, 2022A&A...659A...1S}, the Australian Square Kilometer Array Pathfinder (ASKAP) Evolutionary Map of the Universe (EMU) survey \citep{2011PASA...28..215N}, and polarization observations from the Polarization Sky Survey of the Universe's magnetism (POSSUM) \citep{2010AAS...21547013G} could help detect other similar filaments. 

\begin{acknowledgements}

We acknowledge and thank the referee for their helpful comments that improved this manuscript. This research has made use of the NASA Astrophysics Data System (ADS). This research made use of Astropy,\footnote{http://www.astropy.org} a community-developed core Python package for Astronomy. We acknowledge the use of NASA's \textit{SkyView} facility located at NASA Goddard Space Flight Center.
This research made use of Montage. It is funded by the National Science Foundation under Grant Number ACI-1440620, and was previously funded by the National Aeronautics and Space Administration's Earth Science Technology Office, Computation Technologies Project, under Cooperative Agreement Number NCC5-626 between NASA and the California Institute of Technology.
This research made use of APLpy, an open-source plotting package for Python \citep{aplpy2012,aplpy2019}.
The Dunlap Institute is funded through an endowment established by the David Dunlap family and the University of Toronto. J.L.W. and B.M.G. acknowledge the support of the Natural Sciences and Engineering Research Council of Canada (NSERC) through grant RGPIN-2015-05948, and of the Canada Research Chairs program. S.S.H acknowledges support from NSERC through the Discovery Grants and the Canada Research Chairs programs and by the Canadian Space Agency.  J.L.C. acknowledges support from the Ontario Graduate Student Scholarship. JMS acknowledges the support of the Natural Sciences and Engineering Research Council of Canada (NSERC), 2019-04848.
The DRAO Synthesis Telescope is operated by the National Research Council Canada as a national facility. This paper makes use of data obtained as part of the IGAPS merger of the IPHAS and UVEX surveys carried out at the Isaac Newton Telescope (INT). The INT is operated on the island of La Palma by the Isaac Newton Group in the Spanish Observatorio del Roque de los Muchachos of the Instituto de Astrofisica de Canarias. All IGAPS data were processed by the Cambridge Astronomical Survey Unit, at the Institute of Astronomy in Cambridge. The uniformly-calibrated bandmerged IGAPS catalogue was assembled using the high performance computing cluster via the Centre for Astrophysics Research, University of Hertfordshire.

\end{acknowledgements}

\bibliography{references}{}
\bibliographystyle{aasjournal}

\newpage
\newpage
\clearpage

\end{document}